\documentclass[floats,twocolumn,aps,prb,longbibliography]{revtex4-2}

\usepackage{graphicx}
\usepackage{amsfonts}
\usepackage{amssymb}
\usepackage{amsbsy}
\usepackage{amsmath}
\usepackage[normalem]{ulem}
\usepackage[colorlinks = true,
            linkcolor = magenta,
            urlcolor  = magenta,
            citecolor = magenta,
            anchorcolor = magenta]{hyperref}
\DeclareUnicodeCharacter{2212}{-}
\newcommand{\cdag}{c^\dagger}

\usepackage{subfigure} 
\usepackage{xcolor}



\begin{document}

\title{Spatial extent and entanglement of the Kondo cloud around multiorbital impurities}

\author{A. A. Orlandini}
\affiliation {Instituto de F\'{\i}sica Rosario (CONICET) and Facultad de Ciencias Exactas, Ingeniería y Agrimensura, Universidad Nacional de Rosario, 2000 Rosario, Argentina} 
\author{G. G. Blesio}
\affiliation {Instituto de F\'{\i}sica Rosario (CONICET) and Facultad de Ciencias Exactas, Ingeniería y Agrimensura, Universidad Nacional de Rosario, 2000 Rosario, Argentina} 
\author{C. J. Gazza}
\affiliation {Instituto de F\'{\i}sica Rosario (CONICET) and Facultad de Ciencias Exactas, Ingeniería y Agrimensura, Universidad Nacional de Rosario, 2000 Rosario, Argentina} 
\author{L. O. Manuel}
\affiliation {Instituto de F\'{\i}sica Rosario (CONICET) and Facultad de Ciencias Exactas, Ingeniería y Agrimensura, Universidad Nacional de Rosario, 2000 Rosario, Argentina} 

\begin{abstract}
Although the Kondo effect has been studied for decades, the real-space structure and entanglement of the screening cloud remain largely unexplored for multiorbital impurities with Hund's coupling. Using the density matrix renormalization group (DMRG), we study a two-orbital, two-channel Anderson impurity model in which Hund's coupling stabilizes 
a spin-1 moment. A single-ion anisotropy $D$ drives a topological quantum phase transition between a usual Fermi liquid and a non-Landau Fermi liquid (NLFL). 
In the Kondo phase, as for a spin-1/2 impurity, a single length controls how the screening cloud fits into a finite system.
A natural-orbital analysis reveals two effective screening orbitals, one per channel, whose spatial extent changes across the phase 
diagram, so that deep in each phase the ground state is captured by a few degrees of freedom. The von Neumann entropy between the two channels, each with 
its impurity orbital, takes the values $\ln 2$ for easy-axis anisotropy, where the usual Fermi liquid is reached through a fourth-order Kondo effect, and $\ln 4$ in the NLFL phase, and distinguishes the two Fermi liquids even where the spin
correlations between the two screening orbitals, one per channel, are nearly indistinguishable. Finally, we show that a magnetic field comparable to the anisotropy reactivates the entanglement between the impurity 
and the conduction channels, at the expense of the interchannel entanglement, and that the entanglement of the Kondo singlet is robust against charge fluctuations as 
long as two-electron states dominate the impurity occupancy. These results establish entanglement as a probe of multiorbital Kondo impurities that resolves ground states indistinguishable through their spin correlations.

\end{abstract}

\maketitle 


\section{Introduction}
\label{Sec:Intro}

The Kondo effect, the many-body screening of a localized magnetic moment by the conduction electrons, has evolved from its original explanation of the resistivity minimum in dilute magnetic alloys~\cite{KONDO-PAPER} into a far-reaching paradigm in condensed matter physics~\cite{Georges1996,Kouwenhoven2001} observed in a broad spectrum of physical systems such as magnetic impurities in metals, magnetic atoms and molecules on surfaces~\cite{Madhavan1998,Li1998,Otte2008,Minamitani2012,Spinelli2015,Ormaza2017,Hiraoka2017}, quantum dots~\cite{GoldhaberGordon1998,Cronenwett1998,Pustilnik2001,Potok2007,Iftikhar2015}, and single-molecule junctions~\cite{Liang2002,Parks2010}. The interest in these systems has persisted to the present day, not only due to the large number of materials that exhibit this phenomenology, but also thanks to the various physical systems and experimental tools that have recently emerged~\cite{Evers2020,Yang2019,Verlhac2019}, enabling the study of the Kondo effect from new perspectives~\cite{Borzenets2020,Lee2015,Shim2023}. 
In particular, the screening cloud, long regarded as experimentally elusive, has recently been observed directly in a quantum-dot device~\cite{Borzenets2020}.

The ground state of the single-impurity Kondo model is a nontrivial many-body singlet formed between the 
impurity spin and a fraction of the conduction electrons. This state originates from spin exchange processes 
that lead to the screening of the impurity at temperatures below the Kondo temperature $T_K$. 
Although entanglement is the essence of this phenomenon, its explicit structure has historically received 
little attention. Early analyses of the Kondo ground state showed that the impurity spin is maximally entangled with the conduction electrons as a whole, but not with any single conduction electron, even within the screening length, although the two are spin correlated~\cite{oh2006}.
DMRG studies of the impurity contribution to the entanglement entropy established that it is a 
universal function of $x/\xi_K$, with $x$ the size of the region around the impurity and $\xi_K=\hbar v_F/k_BT_K$, thereby providing a direct measure of the 
size of the screening cloud~\cite{Sorensen2007,Affleck2009}; the same single length was recently shown to govern also the magnetization of part of the system~\cite{Sorensen2024}. The same holds for the 
impurity-conduction electron spin correlations, which in one dimension decay as $1/x$ inside the cloud and as $1/x^2$ outside it~\cite{BARZYKIN-1996,BARZYNKIN-1998,Sorensen96,BORDA-2007,HOLZNER-2009,Busser10}.
In a gapless $J_1$–$J_2$ spin chain with an impurity at one end, which reproduces the low-energy physics of the Kondo model, the ground state is richer than a simple singlet: the Kondo cloud is maximally entangled with the impurity and, at the same time, significantly entangled with the rest of the chain~\cite{bayat2010}. 

In recent years, the internal structure of the Kondo singlet has attracted growing attention. 
Concurrence and von Neumann entropy calculations show that, as the Kondo coupling $J_K$ increases, 
the entanglement concentrates near the impurity in real space while, in energy, it shifts from the Fermi level to high-energy states~\cite{Yang2017}; splitting the bath into high- and low-energy sectors, the low-energy entanglement at Fermi-liquid fixed points flows to parameter-independent values, integer multiples of $\ln 2$~\cite{Zhou2026}. Moreover, in the single-orbital Anderson model the entanglement between opposite-spin electrons is a universal 
monotonic function of the quasiparticle mass enhancement $Z$~\cite{pari2020}, suggesting that 
entanglement could be quantified through a measurable quantity. 
Subsequent works extended these ideas to more complex settings. For a Kondo impurity in a 
pseudogapped bath, the impurity entanglement is a universal function of distance scaled by an entanglement length, which diverges at the Kondo-destruction quantum critical point~\cite{wagner2018}, and even in local-moment phases, where the impurity is not screened, the conduction electrons form a spin cloud of universal form~\cite{Kim2026}.
In multichannel Kondo systems, the impurity–bath entanglement is maximal at zero temperature regardless of the number of channels and decays as a universal power law at low temperature~\cite{kim2021}, 
while its spatial structure forms hierarchical shells~\cite{Shim2023}, and spin-chain realizations have resolved the entanglement structure of the two-channel Kondo model~\cite{Alkurtass2016} and realized multichannel Kondo impurities~\cite{Gaines2026}.
In the two-impurity Kondo model, the critical point between Kondo screening and an interimpurity singlet appears as a change of the partner with which each moment is mainly entangled~\cite{CardenesWuttig2026}. For multilevel impurities, the entanglement entropy and mutual information of one or two impurity sites can be written in terms of local expectation values and correlation functions~\cite{Nishikawa2025}, and impurity--environment correlators identify which mechanism screens the impurity~\cite{Stocker2026}.
In parallel, it has been argued that, once expressed in the basis of natural orbitals, Kondo ground states have an essentially few-body structure: a few natural orbitals suffice in most of the 
parameter space, although genuinely many-body correlations emerge when the Kondo temperature becomes exponentially small, for instance near a quantum critical point~\cite{DEBERTOLIS2021,Zheng2020}.

Multichannel and multiorbital impurity models, however, still pose many open questions. A paradigmatic example is the two-channel, two-orbital Anderson impurity model (2CAM), in which a 
strong Hund’s coupling stabilizes a spin-1 moment on the impurity. In this model, the impurity can be fully screened, giving rise to a fully compensated Kondo effect~\cite{NOZIERES}. 
Despite a much lower Kondo temperature than its spin-1/2 counterpart~\cite{Nevidomskyy2009}, this phase is a Landau Fermi liquid with properties similar to those of the conventional single-channel 
Kondo effect~\cite{Blesio2019}. The richer structure of the impurity, however, gives rise to new phenomena. For instance, a uniaxial single-ion anisotropy acting on the spin-1 moment drives a topological 
quantum phase transition to a non-Landau Fermi liquid~\cite{Blesio2018}, a phase proposed to describe iron phthalocyanine on Au(111)~\cite{Zitko2021}. The transition is not accompanied by any symmetry
breaking; instead, it is signaled by a jump in the Luttinger integral~\cite{Blesio2018}. The 2CAM also describes Ni impurities in O-doped Au chains~\cite{Barral2017,Blesio2018}, Fe atoms 
on MoS$_2$/Au(111)~\cite{BLESIO-ALIGIA-2023}, and nickelocene on Cu(100)~\cite{BLESIO-SCIPOST-2023}; see Ref.~\cite{BLESIO-REVIEW-2024} for a review.

In a recent Letter we showed that the two Fermi liquids of this model have markedly different entanglement structures~\cite{Orlandini2025}. 
In the Kondo phase, the entangled region around the impurity tracks the Kondo cloud; in the non-Landau phase, the quenched impurity mediates a nontrivial entanglement between the two conduction channels, 
and the entanglement entropy of a symmetric bipartition rises to $\ln4$, a value reproduced by a four-spin toy model.

In this work, 
we shift the focus  to the organization of the Kondo state itself. We examine whether the 
screening cloud of a Hund-coupled spin~1 obeys the same finite-size scaling as that of a spin~1/2, and how it behaves as the transition is approached. 
We identify the orbitals that carry the screening, which determine how many degrees of freedom the ground state actually involves and why four spins suffice to reproduce the entanglement of both phases. 
Finally, we test whether the entanglement that tells the two phases apart is a robust property of each phase or depends on the details of the impurity, such as its charge fluctuations or its occupation, and how a magnetic field acting on the impurity reorganizes it.

Using density matrix renormalization group calculations on the 2CAM and on its charge-frozen counterpart, the two-channel, two-orbital Kondo model (2C2OKM), we find that in the Kondo phase the screening length follows a single-length finite-size scaling, as for spin-1/2 impurities(its slow growth as the transition is approached was already reported in Ref.~\cite{Orlandini2025}). Two natural orbitals, one per channel, carry the screening; their spatial extent changes markedly across the phase diagram, and deep in each phase they reduce the problem to a few spins. The entanglement entropy of the symmetric bipartition has size-independent limits, $\ln2$ for easy-axis anisotropy and $\ln4$ in the non-Landau phase, separated by a step that sharpens as the system grows. This entanglement is robust against charge fluctuations as long as two-electron configurations dominate the impurity occupancy, while a magnetic field reorganizes it rather than simply destroying it: in the non-Landau phase, a field comparable to the anisotropy reactivates the entanglement between the impurity and the channels, at the expense of the entanglement between the two channels. The spatial extent of the screening cloud and the entanglement it carries thus emerge as distinct pieces of information, the second distinguishing the non-Landau Fermi liquid from the usual one, including its fourth-order Kondo regime at easy-axis anisotropy ($D<0$), even where the spin correlations between the two screening orbitals hardly change.

The paper is organized as follows. In Sec.~\ref{Sec:model} we introduce the 2CAM and the 2C2OKM and describe the DMRG calculations. Section~\ref{Sec:ResKC} analyzes the integrated spin correlations, the finite-size scaling of the screening length, and the spatial decay of the correlations between the impurity and the conduction electrons as functions of hybridization and anisotropy. In Sec.~\ref{Sec:NatOrb} we characterize the ground state through its natural orbitals and compare it with a four-spin toy model. Section~\ref{Sec:ResES} is devoted to the von Neumann entropy of several bipartitions: we follow it across the transition in both models and study its dependence on hybridization, on a magnetic field applied to the impurity, and on the impurity occupation. Section~\ref{Sec:Conc} summarizes our conclusions. Appendix~\ref{App:Sigma} derives the sum rule behind the integrated correlation function, and Appendix~\ref{App:toy} presents the toy model.
Appendix~\ref{App:Heff} derives the effective exchange between the channels.

\section{Models and method}
\label{Sec:model}

\begin{figure}[!t]
    \begin{center}
        \includegraphics*[width=1\columnwidth]{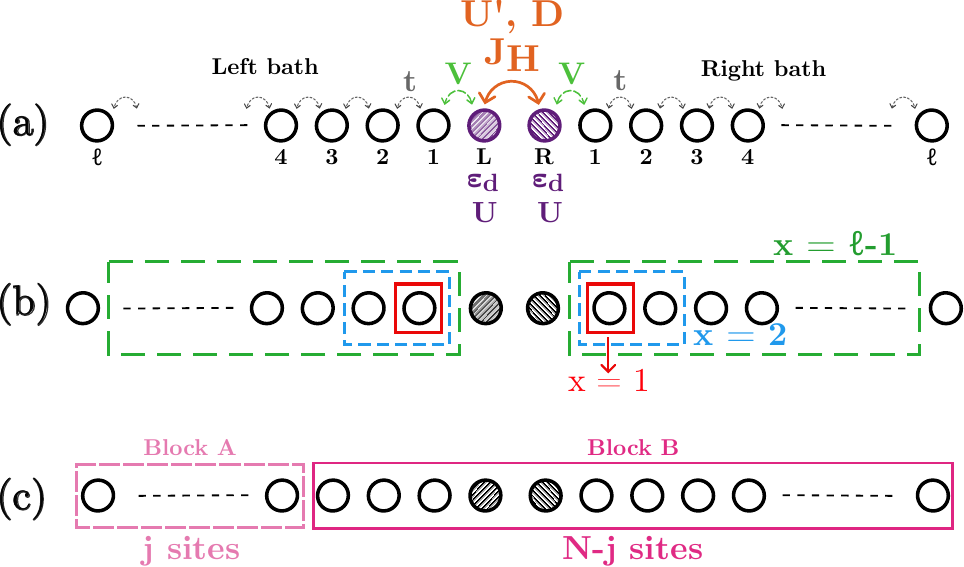}
        \caption{(a) Chain geometry of the 2CAM. In this scheme, impurity orbitals are drawn as dashed circles, while conduction electrons are shown as empty ones. (b) Chain geometry of the 2C2OKM. As in (a), empty circles represent conduction electrons and dashed circles with arrows denote the $s=1/2$ impurity orbitals. (c) Schematic illustration of the spatial coordinate $x$ used in $\Sigma(x)$ for the 2CAM. The variable $x$ is measured symmetrically with respect to the impurities. 
        }
        \label{fig: modelo}
    \end{center}
\end{figure}

We study the two-channel, two-orbital Anderson impurity model (2CAM) with degenerate impurity orbitals, 
each of which, on average, holds approximately one electron. 
A strong Hund coupling $J_H$ induces a spin $S=1$ magnetic moment, while a single-ion anisotropy $D$ favors its $S^z=0$ state, which quenches the moment, for $D>0$, and the doublet $S^z=\pm1$ for $D<0$; we study both signs. Numerical simulations are carried out in a chain geometry, as illustrated in 
Fig.~\ref{fig: modelo}(a). The Hamiltonian of the model is given by
\begin{equation}
    \begin{split}
         H_{2CAM} = H_{band} + H^A_{imp} + H^A_{hyb},
    \end{split}
    \label{ec: Hamiltoniano 2CAM}
\end{equation}
where the conduction channels are modeled as tight-binding chains of length $\ell = \frac{N}{2}-1$, with $N$ the total number of  ``sites'' (conduction channel and impurity orbitals),
\begin{equation}
    H_{band} = -t\sum_{i=1}^{\ell-1}\sum_{\alpha\sigma}  (\cdag_{i{\alpha}\sigma} c_{i+1\alpha\sigma} + {\rm H.c.}),
    \label{ec: HamiltonianoBand}
\end{equation}
and
\begin{equation}
    \begin{split}
        H^A_{imp} &= \sum_\alpha \varepsilon_d n_\alpha + U\sum_\alpha n_{\alpha\uparrow}n_{\alpha\downarrow} + U'n_{L} n_{R}    \\
        & - J_H\mathbf{S}_{L}\cdot\mathbf{S}_{R} +         D\left(S^z_L+S^z_R\right)^2,
    \end{split} 
    \label{ec: HamiltonianoImp}
\end{equation}
\begin{equation}
    H^A_{hyb} = V\sum_{\alpha\sigma}(\cdag_{1\alpha\sigma}d_{\alpha\sigma}+ {\rm H.c.}).
    \label{ec: HamiltonianoHyb}
\end{equation}
In these equations, $d^\dag_{\alpha\sigma}$ ($\cdag_{i{\alpha}\sigma}$) creates an electron on the $\alpha=L,R$ impurity orbital (conduction band) with on-site orbital energy $\varepsilon_d$, and $\mathbf{S}_{\alpha}$ are the spin operators on the $\alpha=L,R$ impurity orbital. We take the hybridization $V$  independent of the energy and the conduction channel.  In Eq. (\ref{ec: HamiltonianoImp}), $U$ is the intraorbital Coulomb repulsion energy, $U'$ corresponds to the interorbital repulsion and $D$ is the single-ion anisotropy. 
We use $\varepsilon_d=-0.4$, $U=J_H=40$, and $U'=10$ unless stated otherwise. The two-electron triplet, 
at energy $2\varepsilon_d+U'-J_H/4=-0.8$, is then the lowest impurity configuration; 
the one-electron configurations lie $0.4$ above it and the empty impurity $0.8$ above it, while 
putting two electrons in the same orbital costs $40$, and three-electron configurations about $60$. The impurity is therefore a spin-1 moment with 
sizable valence fluctuations toward one electron, rather than at the particle-hole symmetric point 
$\varepsilon_d=-U/2-U'$. Since $H^A_{imp}$ contains no pair-hopping term, the number of electrons 
in each channel (a chain plus its impurity orbital) is separately conserved; 
this makes the model genuinely two-channel~\cite{Blesio2018}, and the channels communicate 
only through $U'$, $J_H$, and $D$.

We also present results for a two-channel, two-orbital Kondo impurity model (2C2OKM) described by the Hamiltonian 
\begin{equation}
    H_{2C2OKM} = H_{band} + H^K_{imp} + H^K_{hyb},
    \label{ec: Hamiltoniano 2C2OKM}
\end{equation}
where $H_{band}$ is the same as in Eq. (\ref{ec: HamiltonianoBand}) and
\begin{equation}
    \begin{split}
        H^K_{imp} = - J_H  {\mathbf{S}}_L \cdot {\mathbf{S}}_R + D\left(S^z_L+S^z_R\right)^2, 
    \end{split}
    \label{ec: Hamiltoniano ImpSpin}
\end{equation}
\begin{equation}
        H^K_{hyb} = \sum_\alpha J\hat{\mathbf{S}}_\alpha\cdot\hat{\mathbf{s}}_{1\alpha},
    \label{ec: Hamiltoniano Kondo}
\end{equation}
where $\hat{\mathbf{S}}_\alpha$ ($\alpha=L,R$) are spin-$\tfrac{1}{2}$ operators that replace the two impurity orbitals, each now frozen to single occupancy, $\hat{\mathbf{s}}_{1\alpha}$ are the spin-$\tfrac{1}{2}$ operators for the first conduction site of channel $\alpha$, $J_H$ is the Hund coupling and $J$ is the Kondo coupling. This model allows us to study the system's behavior in the absence of charge fluctuations at the impurity sites. Simulations are also performed in a chain geometry, as shown in Fig.~\ref{fig: modelo}(b). Throughout this work, we will take the 2CAM as the default model. Its name will be omitted unless explicitly comparing with the 2C2OKM. For the 2C2OKM we use $J=0.4$ and $J_H=4$.

To study the spatial properties and entanglement structures of both models, we use the density matrix renormalization group (DMRG)~\cite{DMRG-WHITE} method implemented via the \textit{ITensorMPS} library~\cite{ITENSOR-PAPER}. This numerical method turns out to be the most suitable tool since it solves Eqs. (\ref{ec: Hamiltoniano 2CAM}) and (\ref{ec: Hamiltoniano 2C2OKM}) in a quasi-exact manner, without the need to transform the system to energy space, as the numerical renormalization group does. As a result, the obtained solutions explicitly maintain the spatial structure of the system~\cite{Sorensen96, Affleck10, Busser10, Ribeiro19, Hand06, HOLZNER-2009, Yang2017}. Calculations were performed on systems with up to $L=500$ sites, always taking $N$ as a multiple of four, so that each chain has an odd number of sites $\ell$, with open boundaries, and using a bond dimension
up to $m=800$ to keep the truncation error below $10^{-9}$. 
We conserve the total $S^z$ and, as we neglect the pair-hopping term, we also conserve the number of electrons of each channel. 
The model parameters were chosen to enforce the formation of a spin-$S=1$ impurity moment by taking 
large Hund's coupling values $J_H > W = 1$, where $W$ is the half-bandwidth of the conduction 
channels, so that $t=W/2$ and $v_F=1$ in units of the lattice spacing. In the 2CAM, the hybridization width is $\Delta =  \pi V^2\rho_0(\varepsilon_F)$, where $\rho_0$ is the density 
of states of the conduction baths at the first site of the chain, $\rho_0(\varepsilon_F)=1/(\pi t)$, so that $\Delta=2V^2$.

\textit{Phases of the model.} For $D=0$, the spin-1 moment is fully screened by the two channels and the ground state is an SU(2) 
singlet; this fully compensated Kondo phase is a Landau Fermi liquid~\cite{Blesio2019}. 
For $D<0$, the low-energy impurity doublet $S^z=\pm1$ is connected by $\Delta S^z=2$, so its 
screening requires two spin-flip processes; the effective Kondo coupling is then of second order in 
$J$ (fourth order in $V$), and $\ln T_K\propto-D^2$~\cite{Blesio2019}. We refer to this 
crossover regime, which belongs to the same Fermi-liquid phase, as the fourth-order Kondo 
regime. For $D>0$, the impurity state $S^z=0$ is favored and, at a critical value $D_c\approx2.6\,T_K^0$, with $T_K^0$ the Kondo temperature at $D=0$, the system undergoes a topological quantum phase 
transition into the NLFL~\cite{Blesio2018,Zitko2021}. The Luttinger integral jumps from $I_L=0$ to 
$I_L=\pi/2$, and by the generalized Friedel sum rule, 
$\delta_{\alpha\sigma}=\pi\langle n_{\alpha\sigma}\rangle-I_L$, the phase shift jumps by $\pi/2$ 
at fixed occupancy~\cite{Blesio2018}. Two crossover scales vanish at $D_c$~\cite{Blesio2018},
\begin{align}
    T_K^*(D)&\propto T_K^0\left(\frac{D_c-D}{D_c}\right)^{2}, &(D<D_c),\label{ec:TKstar}\\
    T_q^*(D)&\propto T_K^0\exp\left[-c\left(\frac{T_K^0}{D-D_c}\right)^{1/4}\right], &(D>D_c),\label{ec:Tqstar}
\end{align}
with $c=O(1)$. In the window $T^*\ll T\ll T_K^0$ the impurity behaves as at the two-channel 
Kondo (2CK) fixed point, with a thermodynamic entropy $\tfrac12\ln2$~\cite{COX-ZAWADOWSKI-1998,Blesio2018}. A finite system, which cannot resolve the critical point itself, displays a crossover whose width shrinks with size.

\section{Results and discussion}

\subsection{Kondo cloud and spin correlations}
\label{Sec:ResKC}

In this section we present a spatial characterization of the system based on spin correlations between its electrons. This approach allows us to distinguish between the region responsible for screening the impurity spin (the Kondo cloud) and the rest of the system, providing an analysis of the spatial dependence of the correlations inside both regions. Following Ref.~\cite{HOLZNER-2009}, we define the integrated spin-correlation function for the 2CAM
\begin{equation}
    \Sigma(x)=1+\sum_{i=1}^x \sum_{\alpha=L,R} \frac{\langle \mathbf{S}_{L}\cdot\mathbf{s}_{i\alpha}\rangle}{\langle\mathbf{S}_{L}\cdot\mathbf{S}_{L}\rangle + \langle\mathbf{S}_{R}\cdot\mathbf{S}_{L}\rangle},
    \label{ec: sigma_function}
\end{equation}
where $x$ is the distance to the impurity measured symmetrically along the two conduction chains [see scheme in Fig.~\ref{fig: modelo}(c)], and $\mathbf{s}_{i\alpha}$ is the spin operator of the $i$-th site of the conduction channel $\alpha$. Due to the reflection symmetry of the Hamiltonian~\eqref{ec: Hamiltoniano 2CAM}, the correlations of $\mathbf{S}_R$ with the conduction electrons are the same as those of $\mathbf{S}_L$, and then using one impurity orbital or the other in Eq.~(\ref{ec: sigma_function}) gives the same result. Moreover, this definition is equivalent to the one in terms of the total impurity spin $\mathbf{S}_{imp}=\mathbf{S}_L+\mathbf{S}_R$, since the reflection symmetry gives $\sum_\alpha\langle\mathbf{S}_{imp}\cdot\mathbf{s}_{i\alpha}\rangle=2\sum_\alpha\langle\mathbf{S}_{L}\cdot\mathbf{s}_{i\alpha}\rangle$ and $\langle\mathbf{S}_{imp}\cdot\mathbf{S}_{imp}\rangle=2\left(\langle\mathbf{S}_{L}\cdot\mathbf{S}_{L}\rangle+\langle\mathbf{S}_{R}\cdot\mathbf{S}_{L}\rangle\right)$, so the factor of two cancels between numerator and denominator. This function measures how much the spin of the impurity is screened by the first $x$ sites in each chain; it satisfies the sum rule $\Sigma(\ell)=0$
whenever the ground state is a total spin singlet, i.e., for $D =0$ (Appendix~\ref{App:Sigma}).
In the same way, the screening length is defined as the minimum number of sites per chain 
for which the spin of the impurity has decreased by $90\%$ of its value at the impurity site: 
$\xi_{0.9}(L)=\min\{x;\Sigma(x)<0.1\}$, a criterion used before for the single-impurity Anderson and Kondo models~\cite{Gubernatis1987,Hand06,GAZZA-TORIO-2006,HOLZNER-2009}. Its thermodynamic limit, $L\rightarrow\infty$, is denoted as $\xi$.

\begin{figure}[t!]
    \begin{center}
        \includegraphics*[width=1\columnwidth]{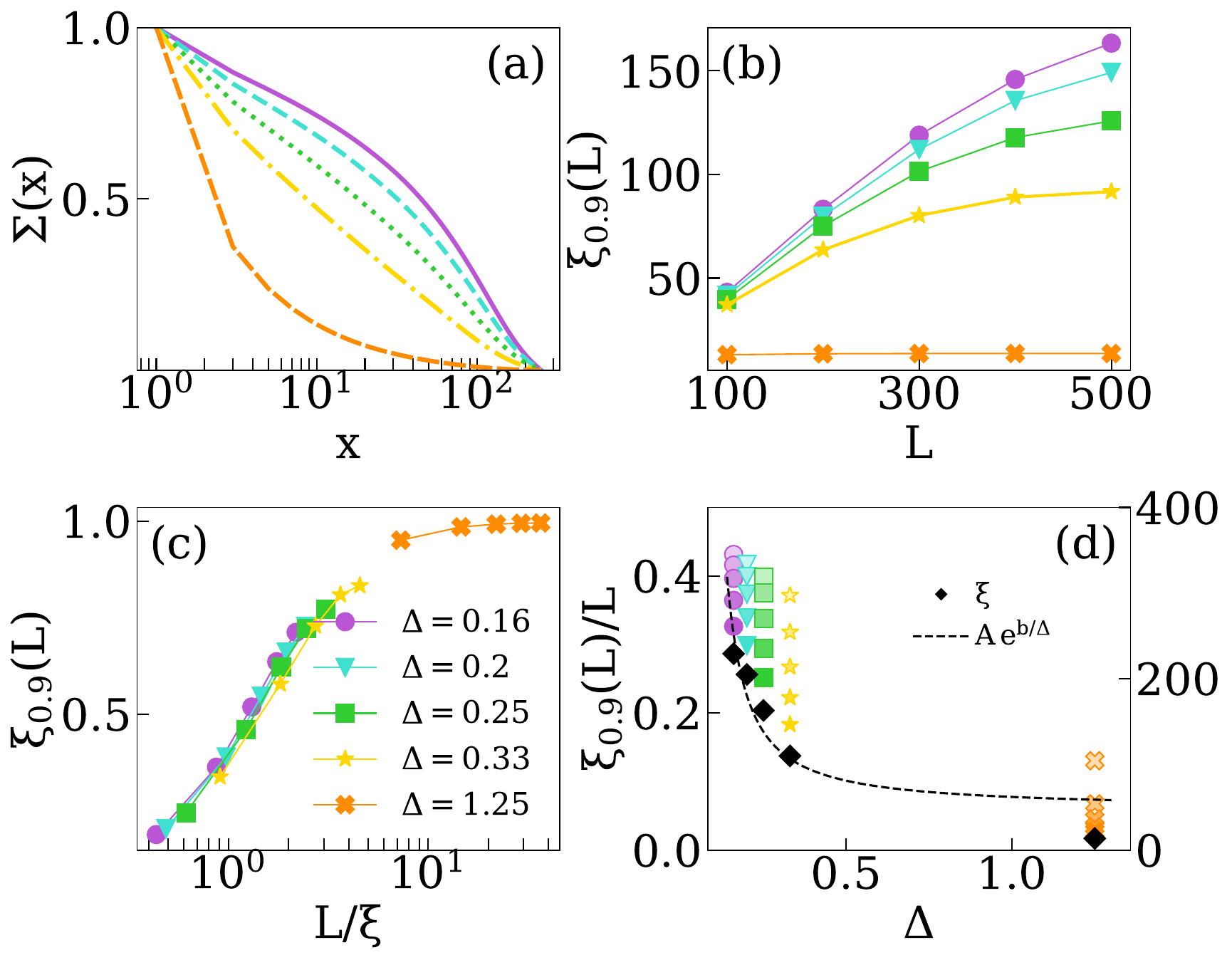}
        \caption{(a) Integrated spin-spin correlations $\Sigma(x)$ as a function of the distance from the impurity $x$, shown for different hybridization values and $L=500$. (b) Screening length $\xi$ as a function of system size $L$. 
        (c) Collapse of $\xi(L)$ under the rescaling $L \rightarrow L/\xi_{0.9}$. (d) Ratio $\xi_{0.9}(L)/L$ as a function of $\Delta$ for $L\in [100, 500]$ (left axis), with color intensity increasing with $L$. Black diamonds show the thermodynamic-limit values $\xi$ (right axis), and the dashed line is  a fit of the form $\xi=A\,e^{\beta/\Delta}$, shown as a guide (see text). This form does not describe the data at large hybridization: for $\Delta\to\infty$ it tends to the constant $A$, of several tens of sites, whereas the cloud should shrink to a few sites, as the point at $\Delta=1.25$ shows. 
        In (b) and (c), points represent numerical data and lines are visual guides. Parameters: $\varepsilon_d = -0.4$, $U = J_H = 40$, $U'=10$, $W=1$, $D=0$. 
        } 
        \label{fig: Colapsos}
    \end{center}
\end{figure} 

One of the most notable features of the Kondo effect is its scaling behavior, meaning that many physical properties of the system follow universal laws when expressed in terms of characteristic energy or length scales. In the case of spatial properties, universality is governed by the size of the Kondo cloud $\xi$, with spatial properties turning into scaling functions of $x/\xi$ rather than depending on these quantities separately~\cite{Sorensen2024}. This universality is also present in more complex Kondo systems, such as the 2CAM. Figures~\ref{fig: Colapsos}(a) and (b) show $\Sigma(x)$ for chains up to $500$ sites and $\xi(L)$, respectively, for different hybridization values. By analyzing $\Sigma(x)$ and $\xi(L)$ across a range of system sizes $N$, we observe that the Kondo screening length $\xi(L)$ becomes largely independent of $N$ for $N \geq 400$ for hybridizations $\Delta>0.2$. Extrapolating the data sets in Fig.~\ref{fig: Colapsos}(b), by means of a power law fit, allows us to determine the screening length in the thermodynamic limit, $\xi$. 
Rescaling the curves in Fig.~\ref{fig: Colapsos}~(b) by means of the Kondo screening length $\xi$ results in all of them collapsing to a single universal behavior curve, as shown in Fig. \ref{fig: Colapsos}~(c), regardless of the parameters of the Hamiltonian. 
Since each curve is rescaled by its own fitted $\xi$, the collapse shows that the 
finite-size crossover is governed by a single length, but it does not by itself 
fix how this length depends on the microscopic parameters.

Although this scaling behavior could have been anticipated from the presence of a fully screened Kondo phase, it is interesting to note that such behavior persists even in the multi-orbital and multi-channel geometry of the 2CAM. Because Hund's coupling makes $T_K^{S=1}\propto(T_K^{S=1/2})^2$~\cite{Nevidomskyy2009,Blesio2019}, multiorbital clouds are parametrically larger than their spin-$1/2$ counterparts at comparable hybridization, which is relevant for spatially resolved probes~\cite{Borzenets2020}.
For $\Delta = 1.25$, however, the system is no longer within the Kondo regime and the universality breaks down. As a result, the rescaled curve deviates significantly from the universal form: there the screening length is only of the order of ten sites and already saturates at the smallest system size ($L = 100$), as seen in Fig.~\ref{fig: Colapsos}(b).

The dependence of the screening length on the hybridization is shown in Fig.~\ref{fig: Colapsos}(d). $\xi_{0.9}$ grows as $\Delta$ decreases, as expected for a Kondo screening cloud. Renormalization-group arguments indicate that the size of the cloud is set by the Kondo temperature~\cite{Sorensen96, Affleck10} $\xi \propto 1/T_K$,
and since $T_K \propto e^{-b/\Delta}$, with $b$ a constant set by the interaction parameters, one expects $\xi \propto e^{b/\Delta}$. A fit of the form $\xi=A\,e^{\beta/\Delta}$ describes the extrapolated values in this range, with $A\simeq 50$ sites and $\beta\simeq 0.3$ [dashed line in Fig.~\ref{fig: Colapsos}(d)]; $\beta$ is an effective exponent for the accessible sizes. The relation $\xi\propto1/T_K$ was verified in this model at fixed $\Delta$ for $D<0$~\cite{Orlandini2025}.

\begin{figure}[t!]
    \begin{center}
        \includegraphics*[width=1\columnwidth]{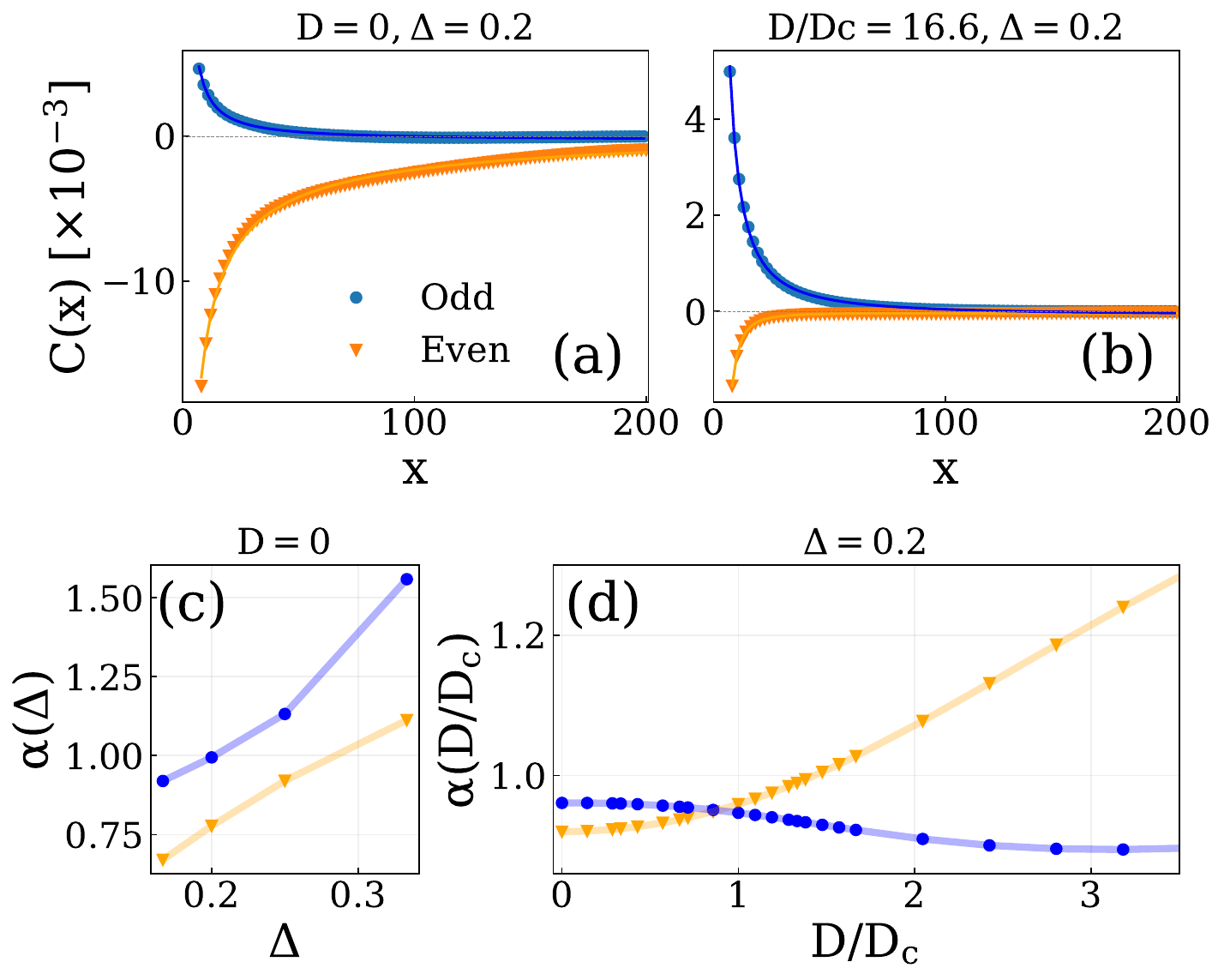}
        \caption{(a) Correlations $C(x)$ as a function of the distance to the impurity for the even (triangles) and odd (circles) subchains in the 2CAM with $\Delta=0.2$, $D=0$ and $L=400$; (b) is the same for $D=0.2$.  Markers represent numerical data and solid lines show the $a+b/x^\alpha$ fits. (c) Fit exponent $\alpha$ for both subchains inside the Kondo cloud as a function of the hybridization $\Delta$, and (d) the single-ion anisotropy $D/D_c$ for $\Delta=0.2$. In these last plots, markers represent numerical data and solid lines serve as a guide for the eye. Parameters: $\varepsilon_d = -0.4$, $U = J_H = 40$, $U'=10$, $W=1$, $D_c\sim0.012$ and $L=400$.
        }
        \label{fig: correlaciones_hyb_anis}
    \end{center}
\end{figure} 

According to perturbation theory, inside the Kondo cloud the correlation between the impurity and the electrons of a one-dimensional conduction band decays as the inverse of the distance from the impurity~\cite{BARZYNKIN-1998}. Figure~\ref{fig: correlaciones_hyb_anis}(a) shows the spin correlations $C(x)$ between the impurity and two consecutive electrons inside the Kondo cloud\footnote{Using $S_L$ or $S_R$ gives the same results because of the symmetry of the Hamiltonian.} 
\begin{equation}
    C(x)=\frac{1}{2}\left(\langle{\hat{\mathbf{S}}_{L}\cdot\hat{\mathbf{s}}_{x_L}}\rangle+\langle{\hat{\mathbf{S}}_{L}\cdot\hat{\mathbf{s}}_{x_R}}\rangle\right),
\end{equation}
where $x$ runs over the chain sites symmetrically with respect to the impurity. Since the conduction bands are half filled, $k_F=\pi/2$ and the spatial dependence of the correlations carries oscillations $\propto\cos(2k_Fx)=(-1)^x$, so the correlations alternate from one site to the next; for this reason the even and odd values of $x$ are plotted separately. These electrons are correlated with the impurity in such a way that this magnitude decreases with distance from it as $1/x$ in one dimension~\cite{BARZYNKIN-1998, Hand06, HAND-TESIS}. For the 2CAM we consider a decay of the form $C(x)=a+b/x^{\alpha}$, whose exponent may depart from the value $\alpha=1$ of the usual single-channel Kondo effect because of the orbital structure of the impurity. Figure~\ref{fig: correlaciones_hyb_anis}(c) shows how the parameter $\alpha$ increases its value with the hybridization, coinciding with the fact that the number of conduction electrons needed to screen the spin of the impurity decreases as the coupling with the baths increases. 

This model undergoes a TQPT for a positive value of anisotropy $D_c$~\cite{Blesio2018}. 
In this case, the ground state is no longer a Kondo singlet between the impurity and the conduction baths. Therefore, we cannot talk about a screening length as in the Kondo phase. 
In this phase, as shown in Figs.~\ref{fig: correlaciones_hyb_anis}(b) and (d), 
it is found that these correlations decay as $1/x^{\alpha_D}$ where $\alpha_D\neq\alpha$
decreases as the anisotropy moves away from its critical value for the odd chain 
(the one with higher correlations) and increases for the even one. 
The power law obtained indicates that although the $S^z=0$ component of the triplet in the impurity 
is strongly favored, it continues to be correlated with many conduction electrons, much as the conduction electrons form a spin cloud around an unscreened moment in local-moment phases~\cite{Kim2026}. 
Therefore, qualitatively, the spin correlation decay is the same within the Kondo 
cloud as in the $D > D_c$ phase. 
However, it is found that the spin correlations decrease appreciably with anisotropy. 
The results obtained indicate that, for the system sizes studied here, there is no abrupt change in the spatial decay of the impurity–conduction electron correlations when the quantum phase transition is crossed, but rather that it changes monotonically as the anisotropy varies.

\begin{figure}[b!]
    \begin{center}
        \includegraphics*[width=1\columnwidth]{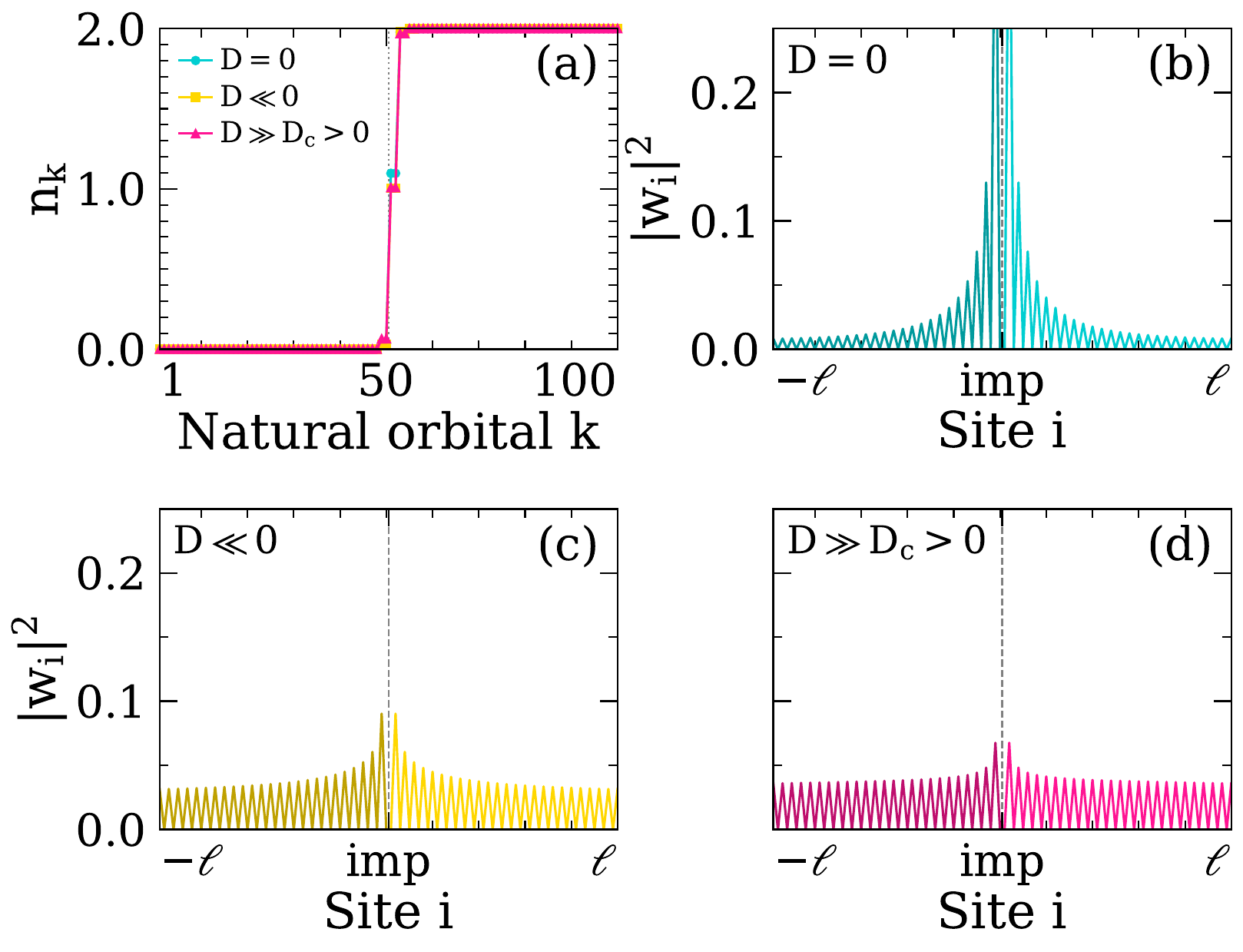}
        \caption{(a) Occupancies $n_\alpha$ of the calculated natural orbitals. Real-space weights $|\text{w}_i|^2$ of the two active natural orbitals for (b) the usual Kondo phase ($D=0$), (c) the fourth-order Kondo regime ($D<0$) and (d) the NLFL phase ($D\gg D_c>0$). Each natural orbital is supported on a single chain: the left (right) half of each panel shows the orbital associated with the left (right) channel. Both orbitals have the same occupation. The two central sites correspond to the first nearest neighbors to the impurity orbitals. Parameters: $\Delta = 0.25$, $\varepsilon_d = -0.5$, $U = 1.0$, $J_H = 1.2$, $U'=0.3$, $W=1$, $D_c\sim0.061$ and $L=100$.
}
        \label{fig: natural_orbitals_extension}
    \end{center}
\end{figure}

\subsection{Natural orbitals}
\label{Sec:NatOrb}
To clarify the structure of the Kondo state we calculate  the natural orbitals (NO) of the system. In the single-impurity Kondo (SIKM) and Anderson models (SIAM), it was shown that the impurity is entangled with a single effective electron whose wave function extends through the chain in real space~\cite{Yang2017,Zheng2020}. In the case of the 2CAM there are expected to be two effective electrons participating in the screening process, giving rise to two dominant natural orbitals. In an interacting system, the natural orbitals are defined as the eigenvectors of the single-particle density matrix 
\begin{equation}
    D_{ij}=\sum_{\sigma} \langle\Psi|c^{\dagger}_{i\sigma}c_{j\sigma}|\Psi\rangle, 
\end{equation}
where $|\Psi\rangle$ is the normalized many-body wave function, $\sigma=\uparrow, \downarrow$ is the spin projection on the $z$ axis and $i$, $j$ refer to the conduction sites (the impurity orbitals are excluded). The occupancies of these effective orbitals are the eigenvalues of the matrix; a natural orbital is considered an \textit{active natural orbital} if it is approximately half-occupied and can, therefore, be correlated with the impurity, otherwise it is called an inactive orbital. 
For an active orbital $f_{\alpha\sigma}=\sum_i w_i\, c_{i\alpha\sigma}$ we define its spin operator as $\mathbf{s}_{NO_\alpha}=\frac12\sum_{\sigma\sigma'}f^\dagger_{\alpha\sigma}\boldsymbol{\sigma}_{\sigma\sigma'}f_{\alpha\sigma'}$, with $\boldsymbol\sigma$ the Pauli matrices.

Since in the 2CAM each impurity orbital is hybridized only with its own conduction channel and we neglect the 
pair-hopping term, the number of electrons in each subsystem (a chain with its corresponding impurity orbital) is conserved separately. As a result, the single-particle density matrix $D_{ij}$ is block-diagonal in the channel index, so each natural orbital is localized on only one of the two chains. There is thus one active natural orbital for the left channel and another for the right channel, both with the same occupancy due to the reflection symmetry of the system [see Fig.~\ref{fig: natural_orbitals_extension}(a)]. In the Kondo regime [see Fig.~\ref{fig: natural_orbitals_extension}(b)], the NO weight is concentrated around the impurity and decays with distance, although for the system size shown it remains finite at the chain end. 

\begin{figure}[t!]
    \begin{center}
        \includegraphics*[width=1\columnwidth]{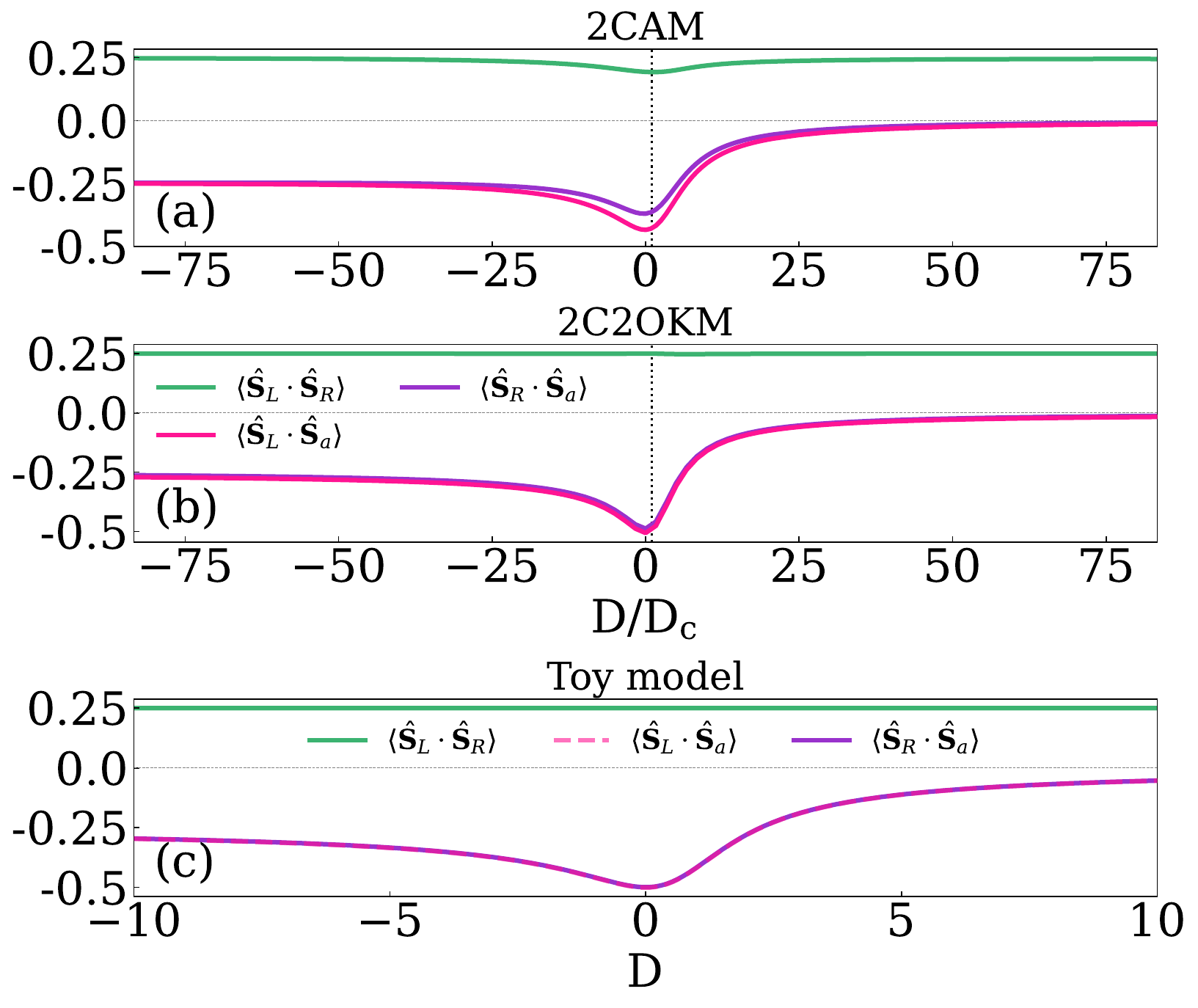}
        \caption{Spin correlations between both natural orbitals (green solid line) and between each of them and the left impurity orbital (pink and purple solid lines) as a function of the single-ion anisotropy $D/D_c$ for the (a) 2CAM with $L=100$, and (b) 2C2OKM with $L=60$. 
        The vertical dashed line indicates the critical value $D=D_c$.
        (c) Spin correlations between pairs of spins in the toy model. 
        Parameters: $\Delta = 0.2$, $\varepsilon_d = -0.4$, $U = J_H = 40$, $U'=10$, $W=1$, $D_c\sim0.012$ and $L=100$ (2CAM); $J=0.4$ and $J_H=4$ (2C2OKM).
        }
        \label{fig: natural_orbitals_correlations}
    \end{center}
\end{figure}

In the fourth-order Kondo regime ($D<-T_K$) [see Fig.~\ref{fig: natural_orbitals_extension}(c)], the NO wave functions remain finite and approximately constant throughout the entire chain, in contrast with the usual Kondo phase. In this region the impurity becomes correlated almost equally with all conduction electrons. This is consistent with the fact that for $D<0$ the Kondo temperature decreases exponentially with $D^2$~\cite{Blesio2019}, and therefore the Kondo cloud ($\xi_K\propto 1/T_K$) becomes larger than the system, giving rise to a more delocalized electron cloud.
The same behavior occurs in the NLFL phase for $D\gg D_c>0$ [see Fig.~\ref{fig: natural_orbitals_extension}(d)]. Although the behaviors seem similar, the nature of their states is not: for $D<0$ a Kondo effect occurs with a very small Kondo temperature, while in contrast for $D\gg D_c>0$ the impurity is quenched and the NO have nothing left to screen. In this phase, no strong correlation between the conduction electrons and the impurity is expected, and indeed the correlation between each natural orbital and the impurity orbital nearly vanishes for $D\gg D_c$ [Fig.~\ref{fig: natural_orbitals_correlations}(a)]. The two natural orbitals remain in a triplet state, as can be seen from the correlation $\left\langle \mathbf s_{NO_L}\cdot\mathbf s_{NO_R}\right\rangle \simeq 1/4$ in Fig.~\ref{fig: natural_orbitals_correlations}(a), which holds for all $D$.

Numerical calculations of the correlations across the phase transition for both models are shown in Figs.~\ref{fig: natural_orbitals_correlations}(a) and (b). From comparing both figures, we find that the dip near $D=0$ in the 2CAM is due to charge fluctuations at the impurity. Similarly, the difference in the correlations of the left impurity orbital with both NO is also related to these fluctuations. Although the correlations between both natural orbitals remain constant along the transition, the ground states at both ends ($D \ll 0$ and $D \gg D_c$) have different nature, as shown in Ref.~\cite{Orlandini2025}, even though in both of them the two natural orbitals are in the $S_T = 1$ subspace. These results are consistent with the baths forming a triplet in the $D\gg D_c$ phase; its $S^z=0$ character follows from the entanglement entropy (Sec.~\ref{Sec:ResES} and Ref.~\cite{Orlandini2025}).

In recent work~\cite{DEBERTOLIS2021} the many-body nature of the ground state in Kondo systems has been questioned, suggesting that it may actually be a few-body state. To explore this in the present 
model, we employed a toy model in which each conduction bath and each impurity orbital is represented by a spin-$1/2$, with a Hamiltonian of the form
\begin{equation}
    \begin{split}
        H_{TM} &= J\left( \hat{\mathbf{s}}_L \cdot \hat{\mathbf{S}}_a + \hat{\mathbf{s}}_R \cdot \hat{\mathbf{S}}_b \right)\\
        &- J_H \hat{\mathbf{S}}_a \cdot \hat{\mathbf{S}}_b + D\left(\hat{S}^z_a+\hat{S}^z_b\right)^2,
        \label{ec: toy}
    \end{split}
\end{equation} 
where $\hat{\mathbf{S}}_a,\; \hat{\mathbf{S}}_b$ correspond to the impurity and $\hat{\mathbf{s}}_L, \hat{\mathbf {s}}_R$ to the left and right conduction channels, respectively. By comparing the 2C2OKM case [see Fig.~\ref{fig: natural_orbitals_correlations}(b)] with the results obtained from the toy model [see Fig.~\ref{fig: natural_orbitals_correlations}(c)], we observe a clear similarity between both cases. In the limit $J_H\to\infty$, the toy model also explains why $\langle \mathbf s_{NO_L}\cdot\mathbf s_{NO_R}\rangle\simeq 1/4$ throughout the phase diagram: once the impurity is projected onto its triplet, the two conduction spins, which belong to different channels and never interact directly, behave as a single spin~1 coupled to the impurity (Appendix~\ref{App:toy}). This suggests that the many-body ground state can be described in terms of a reduced number of effective degrees of freedom. This is consistent with the occupation spectrum of the natural orbitals [see Fig.~\ref{fig: natural_orbitals_extension}(a)], where only two orbitals have occupations appreciably different from 0 or 2. Broadly speaking, the complex interaction between the conduction and impurity degrees of freedom can be captured through a few effective modes, such as the natural orbitals. However, these active degrees of freedom can only be identified after diagonalizing the correlation matrix, which involves solving the original problem with many degrees of freedom. Although the results of the toy model are qualitatively consistent, they are found to be smoother around $D \simeq 0$. Its limited number of degrees of freedom prevents it from capturing the key signals of the phase transition at $D = D_c$: its ground state is analytic in $D$, so it has a crossover at $D\sim J/2$ but no transition (Appendix~\ref{App:toy}), whereas in the extended models the transition shows up in the entanglement entropy (Sec.~\ref{Sec:ResES}). 

These results extend those of Ref.~\cite{DEBERTOLIS2021}, obtained for single-channel Kondo models, to the two-orbital two-channel case: the few-body description captures the ground state deep in both phases, while the transition itself, which involves the gapless continuum, lies beyond it.

\subsection{Entanglement entropy}
\label{Sec:ResES}

In this section we focus on the study of bipartite quantum entanglement in the system. It is natural to approach the Kondo problem from an entanglement perspective, given that the ground state in the conventional phase is a many-body singlet. Moreover, quantum phase transitions, including those not associated with a local order parameter, can be detected through the analysis of entanglement entropy~\cite{WU-2004,CHO-17}. 
As a measure of entanglement, we compute the von Neumann entanglement entropy between two complementary subsystems, $A$ and $B$, defined as $S(A) = -\sum_i\lambda_i\ln\lambda_i$, where $\lambda_i$ are the eigenvalues of the reduced density matrix $\rho$ of subsystem $A$. The impurity cut, which separates the two orbitals from the two channels, is the block of the two central sites; its entropy follows from their reduced density matrix.

In the 2C2OKM at $h=0$, the reduced density matrix of the impurity is diagonal in $M$, the eigenvalue of $\hat S^z_L+\hat S^z_R$, because the total $S^z$ is conserved, and the weights of $M=\pm1$ are equal by spin-flip symmetry. With the impurity projected onto its triplet, the impurity-cut entropy is then a function of the quadrupole $q=\langle(\hat S^z_L+\hat S^z_R)^2\rangle$ alone, as for the spin-one impurity of Ref.~\cite{PIXLEY-2015}; in general, symmetries that make the impurity reduced density matrix diagonal turn its entropy into a function of local expectation values~\cite{Nishikawa2025},
\begin{equation}
    S^{\rm imp}_{e}=-(1-q)\ln(1-q)-q\ln\frac{q}{2}.
    \label{ec:Simpq}
\end{equation}
It equals $\ln2$ for $q=1$ ($D\ll0$), $\ln3$ for $q=2/3$ (isotropic screening), and vanishes for $q\to0$ (NLFL); the value $\ln(2S+1)$ of a fully screened moment holds only with SU(2) symmetry. The channel cut, instead, stays close to $\ln2$ in both phases (exactly $\ln2$ in the toy model, Appendix~\ref{App:toy}): across the transition a channel changes not how much it is entangled but with what, as each impurity changes its main entanglement partner across the critical point of the two-impurity Kondo model~\cite{CardenesWuttig2026}. In the 2CAM both cuts also include the charge fluctuations across the hybridization bond, which the symmetric cut does not cross.

\begin{figure}[t!]
    \begin{center}
        \includegraphics*[width=1\columnwidth]{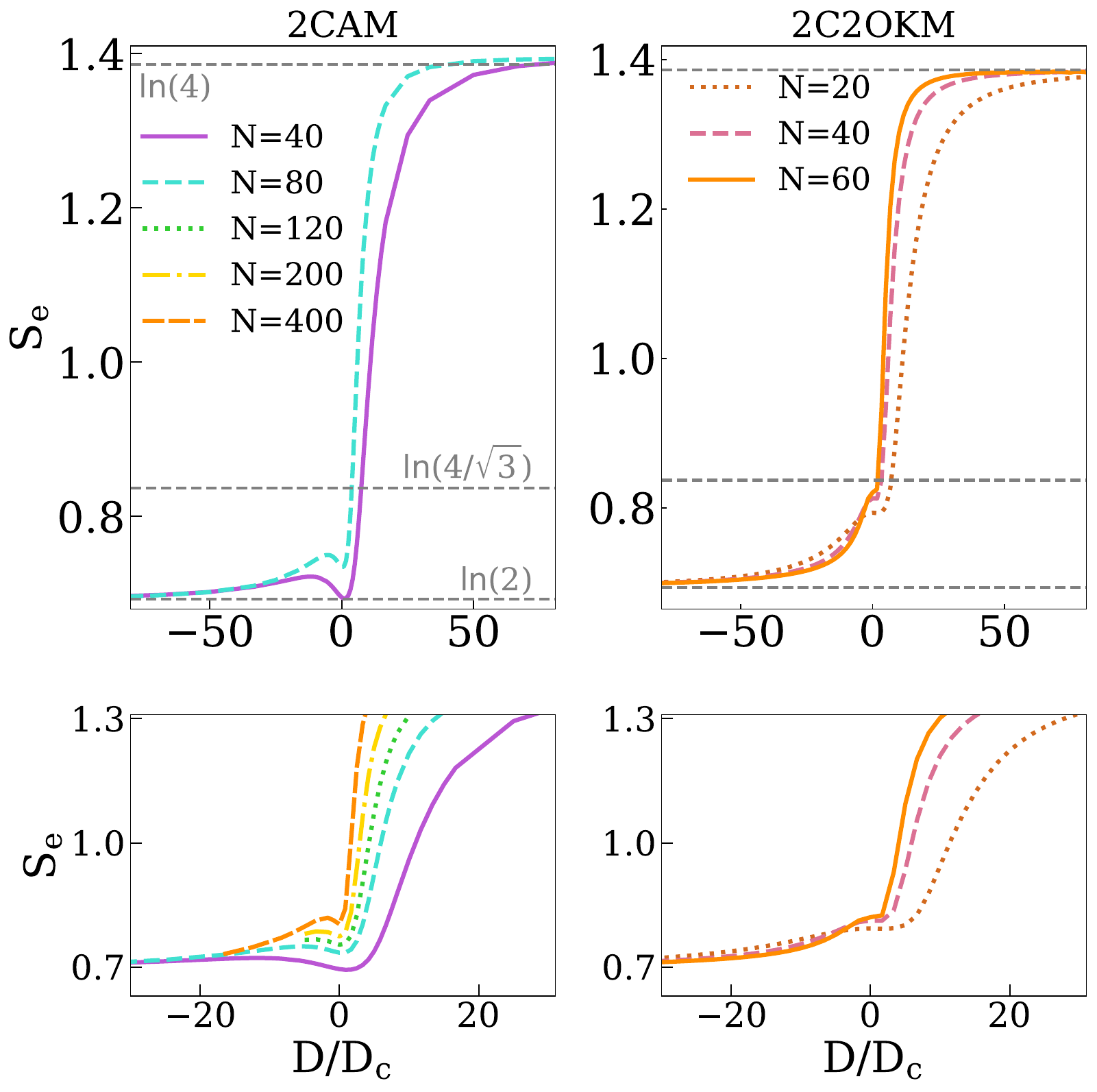}
        \caption{Von Neumann entanglement entropy as a function of the single-ion anisotropy $D/D_c$ dividing the system into two identical subsystems, each consisting of a chain and its corresponding impurity orbital. Panels (a) and (b) show the results for the 2CAM/2C2OKM, respectively, across the full range of anisotropy values, while (c) and (d) provide a zoomed-in view around $D=0$ where a change in the curves is observed as the size of the system increases.  Parameters: $\varepsilon_d=-0.4$, $\Delta=0.2$, $U=J_H=40$, $U'=10$, $D_c\sim0.012$ and $W=1$ (2CAM); $J=0.4$ and $J_H=4$ (2C2OKM).
        }
        \label{fig: entanglement D}
    \end{center}
\end{figure}

\textit{Comparison with the Kondo model—} 
Figs.~\ref{fig: entanglement D}(a) and (b) show how the entanglement entropy evolves when the system is bipartitioned into two identical subsystems, for both models, as a function of the single-ion anisotropy $D$. Roughly speaking, the entropy resembles a step function at $D\approx D_c$ separating two distinct topological phases. In addition, a dip is observed near $D=0$ that shows the crossover between the fourth-order Kondo regime ($D < 0$) and the usual Kondo regime ($0 < D < D_c$). As the system size increases, the rise following the dip becomes steeper, as shown in Fig. \ref{fig: entanglement D}(c) and (d). Extrapolating to the thermodynamic limit, the entanglement entropy is expected to rise from about $\ln(4/\sqrt{3})\simeq 0.84$ at $D=0$, the value of the toy model (Appendix~\ref{App:toy}), and to jump to $\ln 4$ at $D_c$. Since the existence of a TQPT characterized by a finite critical anisotropy $D_c$ has been previously demonstrated~\cite{Blesio2018}, these two limit values can therefore be used to characterize the distinct phases: the NLFL phase ($D > D_c$), associated with an entanglement entropy $S_e^{nL} = \ln(4)$ that is independent of anisotropy, and the conventional Fermi liquid phase ($0 < D < D_c$), where $S_e$ does depend on the anisotropy. In this regime, the value $S_e(D=0)$ increases with the system size for the 2CAM, as occurs for the 2C2OKM. For $D < 0$, the entanglement entropy takes the value $\ln(2)$, which is in correspondence with Numerical Renormalization Group (NRG) calculations~\cite{Blesio2019,GERMAN-TESIS}, which show that in this regime the impurity is Kondo screened and the ground state is a local Fermi liquid. The entropy value $\ln(2)$ corresponds to the state $(|{\Uparrow}\rangle_{imp}|{\downarrow\downarrow}\rangle+|{\Downarrow}\rangle_{imp}|{\uparrow\uparrow}\rangle)/\sqrt2$, with $|{\Uparrow}\rangle$, $|{\Downarrow}\rangle$ the impurity states $S^z=\pm1$ and the arrows on the right the screening electrons of the two channels; it has Schmidt rank two across the cut but is not an SU(2) singlet, since total spin is not conserved for $D\neq0$.
For large positive anisotropy values, the entropy saturates at $\ln(4)$, corresponding to a product state of two $S_z = 0$ triplets: one for the impurity and the other for the two conduction channels. The behavior of the entanglement entropy at these two limits becomes independent of the system size, reminiscent of the parameter-independent multiples of $\ln2$ found at Fermi-liquid fixed points for an energy-space bipartition of the bath~\cite{Zhou2026}. 
However, comparing both sides of Fig.~\ref{fig: entanglement D}, we find that 
the transition around $D\sim D_c$ in the two models becomes increasingly similar as the system sizes increase.

The entanglement entropy together with the toy model allows us to clarify the results obtained from the correlations between both natural orbitals. Although the correlation between the two remains essentially constant throughout the entire transition the ground state at both extremes of the transition ($D \ll 0$ and $D \gg D_c$) has a completely different nature, even though both belong to the subspace with total spin $S_T = 1$. In the first case, the ground state has no entanglement between the conduction baths because the two screening orbitals are $|\uparrow \uparrow\rangle$ or $|\downarrow\downarrow\rangle$, a classical mixture once the impurity is traced out, while in the latter there is a triplet state with $m=0$, between them. This highlights the relevance of entanglement based analyses for distinguishing states that spin correlations on their own fail to differentiate.

The two channels never interact directly, and the quenched impurity cannot couple them at first order: in $|T_0\rangle=(|{\uparrow\downarrow}\rangle+|{\downarrow\uparrow}\rangle)/\sqrt2$, the $M=0$ member of the triplet, every $\langle T_0|\hat{\mathbf S}_\alpha|T_0\rangle$ vanishes. At second order, virtual excursions to $M=\pm1$, at energy $D$, and to the interorbital singlet, at energy $J_H$, generate an interaction between the first sites of the two channels, valid deep in the NLFL phase, $D\gg J$ (Appendix~\ref{App:Heff}),
\begin{equation}
    H_{\rm eff}=-\frac{J^2}{2D}\left(\hat s^x_{1L}\hat s^x_{1R}+\hat s^y_{1L}\hat s^y_{1R}\right)+\frac{J^2}{2J_H}\,\hat s^z_{1L}\hat s^z_{1R}.
    \label{ec:Heff}
\end{equation}
The first term is a ferromagnetic $XY$ exchange: $\hat S^\pm_\alpha$ takes $|T_0\rangle$ to $|T_\pm\rangle$ whichever orbital acts, so a spin flip in one channel can be undone by the opposite flip in the other. Its ground state for two spins $1/2$ is the maximally entangled state $|T_0\rangle_{ch}$, which supplies the second $\ln2$ of the symmetric cut. Since each chain has an odd number of electrons and thus carries a spin $1/2$, $H_{\rm eff}$ selects $|T_0\rangle_{ch}$ however weak it is, so this $\ln2$ does not depend on size. Unlike the two-impurity Kondo model, where a singlet between the impurities couples their baths antiferromagnetically~\cite{bayat2012,BAYAT-2017}, here the mediator is the $M=0$ member of a triplet, the mediated exchange is ferromagnetic and uniaxial, and the channels end up in a copy of the impurity's own state.

\begin{figure}[t!]
    \begin{center}
        \includegraphics*[width=1\columnwidth]{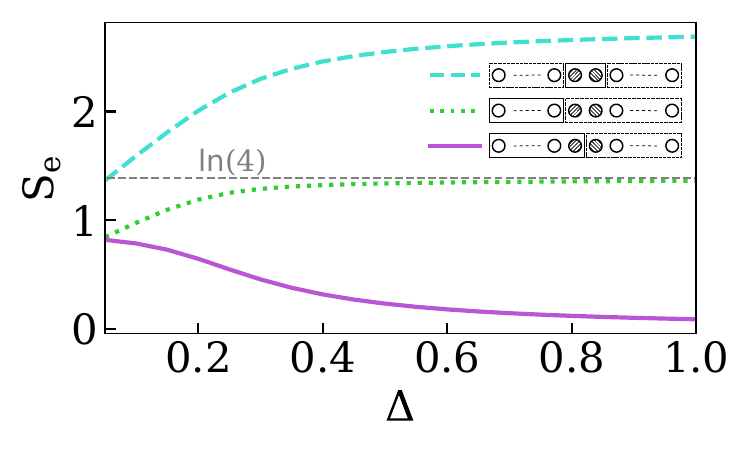}
        \caption{Von Neumann entanglement entropy as a function of the hybridization for three different partitions of the system. Parameters: $\varepsilon_d=-0.4$, $U=J_H=1$, $U'=0$, $D=0$, $W=1$ and $L=300$. 
        }
        \label{fig: entanglement hyb}
    \end{center}
\end{figure}

\begin{figure}[t!]
    \begin{center}
        \includegraphics*[width=1\columnwidth]{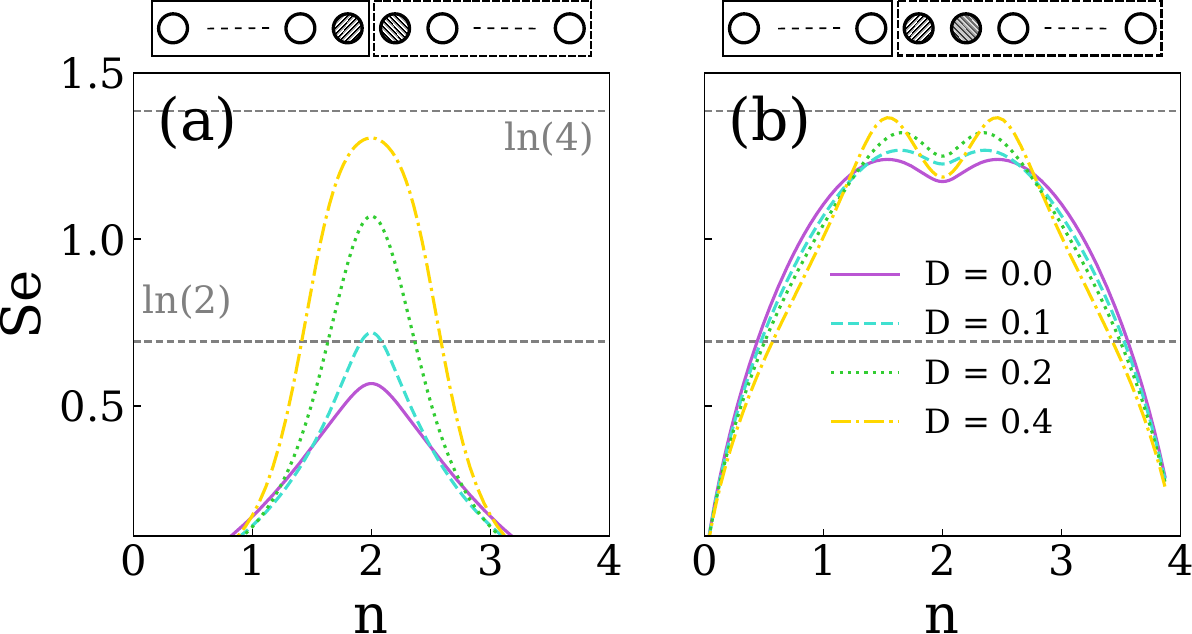}
        \caption{Von Neumann entanglement entropy as a function of the occupancy of the impurity for different values of the single-ion anisotropy $D$. Parameters:  $\Delta=0.2$, $U=J_H=40$, $U'=10$, $W=1$ and $L=40$. 
        }
        \label{fig: entanglement occup}
    \end{center}
\end{figure}

\textit{Hybridization—} 
A deeper understanding of the 2CAM can be achieved by analyzing the dependence of the entanglement entropy on hybridization, as shown in  Fig.~\ref{fig: entanglement hyb}. When the system is bipartitioned symmetrically [see bottom bipartition sketch in Fig.~\ref{fig: entanglement hyb}], the entropy tends to zero as hybridization increases, since the hybridization becomes of the order of $J_H$ and $U$, causing the ground state of the system to consist of the product state of the ground states of each subsystem separately. When a channel is separated from the rest of the system [see middle bipartition sketch in Fig.~\ref{fig: entanglement hyb}], the entropy tends to $\ln(4)$ as hybridization increases. This limit arises from a different type of entanglement than the one that generates the results for $D\gg D_c$ in Fig.~\ref{fig: entanglement D}: at large hybridization, strong charge fluctuations appear in the impurity and the ground state has important contributions from the four states: empty, singly occupied \textit{up}, singly occupied \textit{down} and doubly occupied [see Fig.~\ref{fig: entanglement occup} and the discussion below]. In this way, the limit $\ln(4)$ is obtained from an entanglement between the channel and its respective orbital. Indeed, at large hybridization each impurity orbital forms a doubly occupied bonding orbital with the first site of its chain; each spin species then occupies a mode shared by both sides of the cut and contributes $\ln2$.

Indeed, studying the entanglement between the impurities and the baths (see top bipartition sketch in Fig.~\ref{fig: entanglement hyb}) shows that the entropy tends to the limit $2\ln4=\ln16\simeq2.77$, two independent bonding orbitals each contributing $\ln4$, when hybridization increases. These results provide evidence for the presence of two subsystems, each consisting of an impurity coupled to a single conduction channel. Finally, for small hybridization ($\Delta \rightarrow 0$), the entropy between the impurity and the baths remains close to $2\ln 2$, larger than the value $\ln3$ of a spin $S=1$ screened at $D=0$ [Eq.~(\ref{ec:Simpq}) with $q=2/3$] because charge fluctuations at the impurity contribute significantly. 

\begin{figure}[b!]
    \begin{center}
        \includegraphics*[width=1\columnwidth]{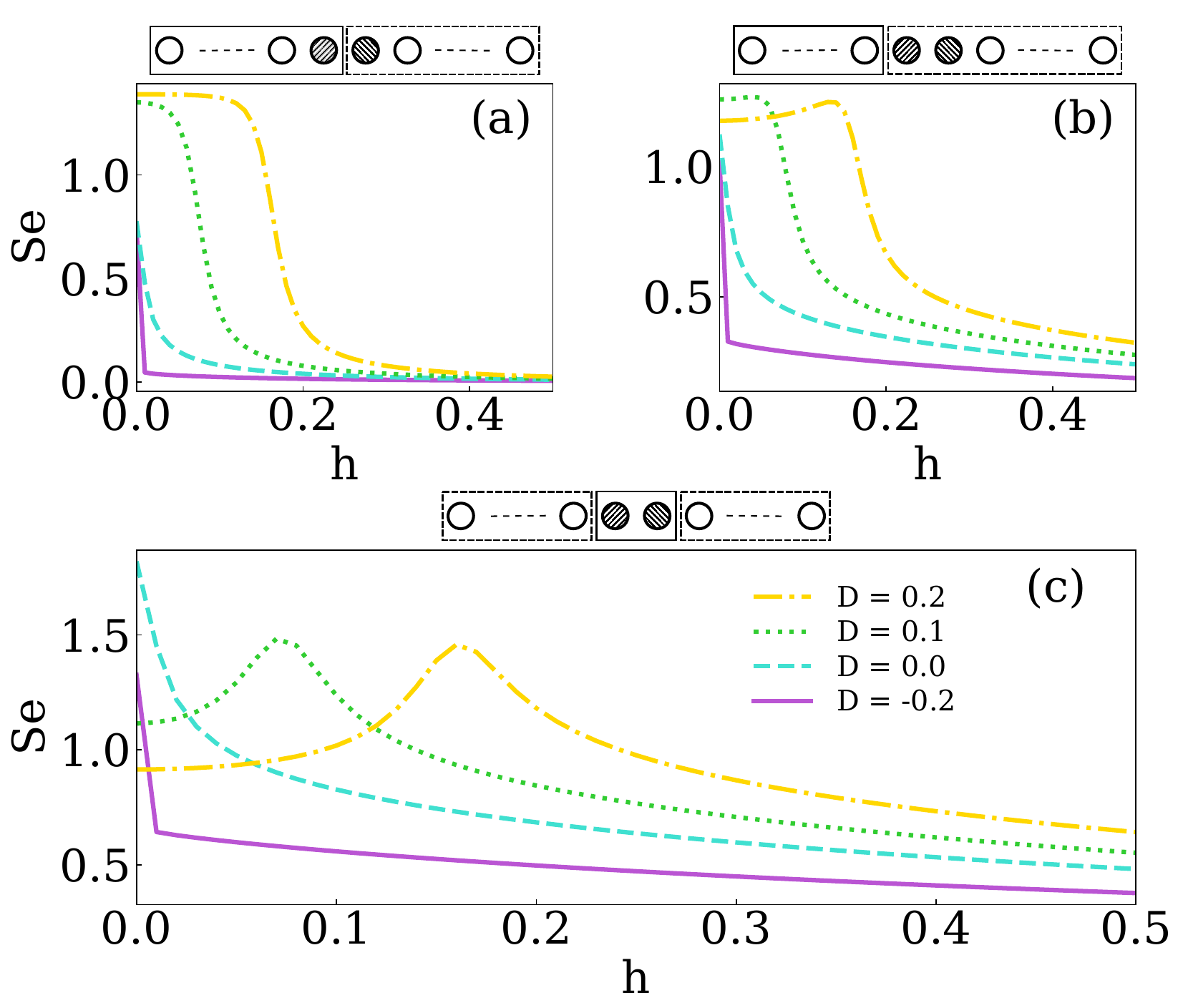}
        \caption{Von Neumann entanglement entropy as a function of the magnetic field $h$ applied to the impurity for various values of the anisotropy $D$. We consider three types of bipartitions: (a) the system divided into two identical subsystems, (b) with one channel separated from the rest and (c) the impurity with both conduction channels. Parameters: $\varepsilon_d=-0.4$, $\Delta=0.2$, $U=J_H=40$, $U'=10$, $W=1$, and $L=200$. }
        \label{fig: entanglement field}
    \end{center}
\end{figure}

\textit{Magnetic field—} 
When a magnetic impurity in a metal is exposed to magnetic fields, the Kondo effect breaks down as the field favors a particular spin projection on the impurity over the others, thus reducing the spin fluctuations necessary for the singlet formation. This leads to a reduction in the entanglement between the impurity and the rest of the system~\cite{Stocker2026}. Here we consider a magnetic field applied only to the impurity, 
in the direction of the single-ion anisotropy axis,  $-h(S^z_L+S^z_R)$, so that $h > 0$ favors $S^z=+1$. We find that the entanglement between the symmetrical subsystems [Fig.~\ref{fig: entanglement field}(a)] decreases much more rapidly with increasing magnetic field than in the case of entanglement between a conduction channel and the rest of the system [Fig.~\ref{fig: entanglement field}(b)]. Thus, a given magnetic field is more effective at disentangling the subsystems than it is at disentangling a channel from its corresponding orbital. This happens because the magnetic field, besides suppressing the Kondo screening, reduces the contribution of the entangled component with $S_z=0$ of the impurity and simultaneously favors the non-entangled projection with $S_z=1$. In general, the magnetic field necessary to destroy a Kondo singlet is of the order of the Kondo temperature $T_K$ in the absence of anisotropy. Since the Kondo temperature for a spin-$1$ impurity is significantly lower than that of a spin-$1/2$ system, it follows that a smaller magnetic field is sufficient to break down the Kondo effect in the 2CAM. 

In the NLFL phase, the single-ion anisotropy $D$ favors a spin projection $S_z=0$, which causes the magnetic field required to disentangle parts of the system to be greater compared to the case when $D<0$. While the anisotropy penalizes the $S_z=\pm 1$ projections relative to $S_z=0$, the presence of a magnetic field lowers the energy of the $S_z=1$ state. When the magnetic field applied to the impurity becomes comparable to the anisotropy, the competition between the two contributions balances out and the two spin projections $|\uparrow_L\uparrow_R\rangle$ and $(|\downarrow_L\uparrow_R\rangle + |\uparrow_L\downarrow_R\rangle)/\sqrt{2}$
become degenerate. As a consequence, a magnetic degree of freedom is reactivated on the impurity, enabling it 
to become entangled again with the conduction channels.
A signature of this is already visible in Fig.~\ref{fig: entanglement field}(b), and it becomes clear in Fig.~\ref{fig: entanglement field}(c): for $D = 0.1$ and $0.2$ the entanglement between the impurity and the conduction baths increases up to $h \sim D$ and decreases again beyond it. Once the magnetic field exceeds the anisotropy it favors specific spin projections, thus breaking the reemergent impurity--channel entanglement and causing the different parts of the system to disentangle. Thus, owing to the multiorbital nature of the impurity, the effect of the magnetic field is more complex than in the usual single-orbital case. While conventional Kondo phases remain fragile in the presence of a magnetic field, other phases such as the NLFL phase need a field of the order of $D$ to be disentangled, and the magnetic field can even induce entangled states between the impurity and the conduction channels.

\textit{Occupation—} 
For symmetric parameters that produce a reflection symmetry around half-filling of the impurity, the system places itself at the symmetric point where particle-hole symmetry holds, thus having an equal contribution of states with one and three electrons. Away from this point, the symmetry is broken and effects associated with the charge transfer begin to emerge. When the von Neumann entanglement entropy is measured by separating the system into two equal subsystems [Fig.~\ref{fig: entanglement occup}(a)], it reaches a maximum at half-filling. At this point, a strong dependence on the anisotropy is observed: $S_e\simeq0.56$ for $D=0$, below $\ln2$, and it grows with $D$ up to $S_e\simeq1.30$ for $D=0.4$, close to $\ln4$.  For anisotropies small compared with $U'$ and $J_H$, the entanglement entropy becomes independent of $D$ when the average total impurity occupancy falls outside the range $1\le n_{imp}\le 3$: approximately two electrons are required so that the anisotropy has an effect on the impurity, because $(S^z_L+S^z_R)^2$ is constant in the one- and three-electron sectors. The entanglement between a conduction channel and the rest of the system [Fig.~\ref{fig: entanglement occup}(b)] depends much more weakly on the anisotropy than in the previous case. Unlike in Fig.~\ref{fig: entanglement occup}(a), in this case the entanglement entropy exhibits two maxima, near $n_{imp} \approx 1.5$ and $2.5$, separated by a local minimum at $n_{imp} = 2$. Increasing $D$ enhances these maxima, which approach $\ln 4$ for $D = 0.4$, whereas at $n_{imp} = 2$ the dependence on $D$ is non-monotonic. For $D\sim 0$, the entanglement entropy varies only weakly over the range of total occupancy of the impurity $1\le n_{imp}\le 3$. This indicates that the entanglement structure of the Kondo singlet in the two-orbital model is robust against fluctuations in the occupancy, as long as there are significant contributions from two-electron states.

\section{Summary and Conclusions}
\label{Sec:Conc}

In this work we have studied the spatial properties of the Anderson model for a magnetic impurity with spin $S=1$, having two orbitals for the impurity coupled to two conduction channels (2CAM), using DMRG. This system has previously been studied using the NRG method~\cite{Blesio2018,Blesio2019}, in which the spatial structure of the conduction bath is not directly accessible. In contrast, DMRG allows us to obtain the spatial properties of the system in a more natural way. We found that in the usual Kondo phase of the 2CAM, the spatial properties already known for simpler models, such as the SIAM, are recovered. This suggests that the properties of the Kondo singlet in the fully screened Kondo regime are qualitatively independent of the orbital structure of the impurity. In the presence of magnetic anisotropy we find that the ground state can be effectively described in terms of a reduced number of degrees of freedom, deep in each phase,  although the sharp transition between the two phases requires the full continuum of the channels. 

By means of the entanglement entropy we have characterized the two phases involved in the topological transition. In the NLFL phase the ground state of the system develops a nontrivial entanglement structure, which is a distinctive hallmark of the phase~\cite{Orlandini2025}. This entanglement can be traced to the effective exchange between the channels mediated by the quenched impurity, Eq.~(\ref{ec:Heff}), a second-order ferromagnetic $XY$ coupling that favors the entangled state $|T_0\rangle_{ch}$. The combined effect of anisotropy and an external magnetic field applied to the impurity revealed the reemergence of the entanglement between the impurity and the channels: the competition between these two contributions balances out in the region $D\sim h$, where the entangled state $|T_0\rangle_{ch}$ of the two channels is destroyed. Moreover, we found that the entanglement of the Kondo singlet is robust against charge fluctuations, remaining nearly unchanged as long as two-electron configurations dominate the impurity occupancy.

\begin{acknowledgments}
\label{Sec:Ack}
We acknowledge financial support by CONICET (Argentina) under Grant PIP No. 3220. Part of the results presented in this work have been obtained by using the facilities of the CCT-Rosario Computational Center, member of the High Performance Computing National System (SNCAD, MinCyT- Argentina).
We also gratefully acknowledge the HPC RIVR consortium (\href{https://www.hpc-rivr.si}{www.hpc-rivr.si}) and EuroHPC JU (\href{https://eurohpc-ju.europa.eu/}{eurohpc-ju.europa.eu}) for funding this research by providing computing resources of the HPC system Vega at the Institute of Information Science (\href{https://www.izum.si/en/home/}{www.izum.si}).
\end{acknowledgments}

\section*{Data availability}
The data that support the findings of this article are available from the authors upon reasonable request.

\bibliography{biblio.bib}

@article{oh2006,
  title = {Entanglement of an impurity and conduction spins in the {K}ondo model},
  author = {Oh, Sangchul and Kim, Jaewan},
  journal = {Phys. Rev. B},
  volume = {73},
  issue = {5},
  pages = {052407},
  numpages = {4},
  year = {2006},
  month = {Feb},
  publisher = {American Physical Society},
  doi = {10.1103/PhysRevB.73.052407},
  url = {https://link.aps.org/doi/10.1103/PhysRevB.73.052407}
}

@article{PIXLEY-2015,
  title = {Entanglement entropy near {K}ondo-destruction quantum critical points},
  author = {Pixley, J. H. and Chowdhury, Tathagata and Miecnikowski, M. T. and Stephens, Jaimie and Wagner, Christopher and Ingersent, Kevin},
  journal = {Phys. Rev. B},
  volume = {91},
  pages = {245122},
  year = {2015},
  doi = {10.1103/PhysRevB.91.245122},
}

@article{BAYAT-2017,
  title = {Scaling of tripartite entanglement at impurity quantum phase transitions},
  author = {Bayat, Abolfazl},
  journal = {Phys. Rev. Lett.},
  volume = {118},
  pages = {036102},
  year = {2017},
  doi = {10.1103/PhysRevLett.118.036102},
}

@article{pari2020,
  title = {Quasiparticle Mass Enhancement as a Measure of Entanglement in the {K}ondo Problem},
  author = {Pari, Nayra A. \'Alvarez and Garc\'{\i}a, D. J. and Cornaglia, Pablo S.},
  journal = {Phys. Rev. Lett.},
  volume = {125},
  issue = {21},
  pages = {217601},
  numpages = {5},
  year = {2020},
  month = {Nov},
  publisher = {American Physical Society},
  doi = {10.1103/PhysRevLett.125.217601},
  url = {https://link.aps.org/doi/10.1103/PhysRevLett.125.217601}
}

@article{bayat2010,
  title = {Negativity as the entanglement measure to probe the {K}ondo regime in the spin-chain {K}ondo model},
  author = {Bayat, Abolfazl and Sodano, Pasquale and Bose, Sougato},
  journal = {Phys. Rev. B},
  volume = {81},
  issue = {6},
  pages = {064429},
  numpages = {5},
  year = {2010},
  month = {Feb},
  publisher = {American Physical Society},
  doi = {10.1103/PhysRevB.81.064429},
  url = {https://link.aps.org/doi/10.1103/PhysRevB.81.064429}
}

@article{bayat2012,
 author = {Bayat, Abolfazl and Bose, Sougato and Sodano, Pasquale and Johannesson, Henrik},
 doi = {10.1103/physrevlett.109.066403},
 journal = {Phys. Rev. Lett.},
 note = {https://arxiv.org/pdf/1201.6668},
 number = {6},
 pages = {066403},
 title = {Entanglement Probe of Two-Impurity {K}ondo Physics in a Spin Chain},
 url = {https://app.dimensions.ai/details/publication/pub.1011526133},
 volume = {109},
 year = {2012}
}

@article{Lee2015,
 author = {Lee, S-S B and Park, Jinhong and Sim, H-S},
 doi = {10.1103/physrevlett.114.057203},
 journal = {Phys. Rev. Lett.},
 number = {5},
 pages = {057203},
 title = {Macroscopic Quantum Entanglement of a {K}ondo Cloud at Finite Temperature},
 url = {https://app.dimensions.ai/details/publication/pub.1043107508},
 volume = {114},
 year = {2015}
}

@article{Yang2017,
  title = {Unveiling the internal entanglement structure of the {K}ondo singlet},
  author = {Yang, Chun and Feiguin, Adrian E.},
  journal = {Phys. Rev. B},
  volume = {95},
  issue = {11},
  pages = {115106},
  numpages = {7},
  year = {2017},
  month = {Mar},
  publisher = {American Physical Society},
  doi = {10.1103/PhysRevB.95.115106},
  url = {https://link.aps.org/doi/10.1103/PhysRevB.95.115106}
}

@article{wagner2018,
 author = {Wagner, Christopher and Chowdhury, Tathagata and Pixley, J H and Ingersent, Kevin},
 doi = {10.1103/physrevlett.121.147602},
 journal = {Phys. Rev. Lett.},
 number = {14},
 pages = {147602},
 title = {Long-Range Entanglement near a {K}ondo-Destruction Quantum Critical Point},
 volume = {121},
 year = {2018}
}

@article{kim2021,
 author = {Kim, Donghoon and Shim, Jeongmin and Sim, H-S},
 doi = {10.1103/physrevlett.127.226801},
 journal = {Phys. Rev. Lett.},
 number = {22},
 pages = {226801},
 title = {Universal Thermal Entanglement of Multichannel {K}ondo Effects},
 url = {https://app.dimensions.ai/details/publication/pub.1142734693},
 volume = {127},
 year = {2021}
}

@article{Shim2023,
 author = {Shim, Jeongmin and Kim, Donghoon and Sim, H.-S.},
 doi = {10.1038/s41467-023-39234-6},
 journal = {Nat. Commun.},
 number = {1},
 pages = {3521},
 title = {Hierarchical entanglement shells of multichannel {K}ondo clouds},
 url = {https://app.dimensions.ai/details/publication/pub.1159788634},
 volume = {14},
 year = {2023}
}

@article{Zheng2020,
    author={Zheng, Ru
    and He, RongQiang
    and Lu, ZhongYi},
    title={Natural orbitals renormalization group approach to a {K}ondo singlet},
    journal={Sci. China Phys. Mech. Astron.},
    year={2020},
    month={May},
    day={07},
    volume={63},
    number={9},
    pages={297411},
    issn={1869-1927},
    doi={10.1007/s11433-019-1520-3},
    url={https://doi.org/10.1007/s11433-019-1520-3}
}

@article{HOLZNER-2009,
   title={Kondo screening cloud in the single-impurity {A}nderson model: A density matrix renormalization group study},
   volume={80},
   ISSN={1550-235X},
   url={http://dx.doi.org/10.1103/PhysRevB.80.205114},
   DOI={10.1103/physrevb.80.205114},
   number={20},
   pages = {205114},
  numpages = {8},
   journal={Phys. Rev. B},
   publisher={American Physical Society (APS)},
   author={Holzner, Andreas and McCulloch, Ian P. and Schollwöck, Ulrich and von Delft, Jan and Heidrich-Meisner, Fabian},
   year={2009},
   month={Nov} }

@article{ITENSOR-PAPER,
  title   = {The {ITensor} Software Library for Tensor Network Calculations},
  author  = {Fishman, Matthew and White, Steven R. and Stoudenmire, E. Miles},
  journal = {SciPost Phys. Codebases}, pages = {4}, year = {2022},
  doi     = {10.21468/SciPostPhysCodeb.4}}

@article{Zitko2021,
   title={Iron phthalocyanine on {A}u(111) is a ``non-{L}andau'' {F}ermi liquid},
   volume={12},
   ISSN={2041-1723},
   url={http://dx.doi.org/10.1038/s41467-021-26339-z},
   DOI={10.1038/s41467-021-26339-z},
   number={1},
   pages = {6027},
   journal={Nat. Commun.},
   publisher={Springer Science and Business Media LLC},
   author={Žitko, R. and Blesio, G. G. and Manuel, L. O. and Aligia, A. A.},
   year={2021},
   month={Oct} }

@article{Blesio2019,
   title={Fully compensated {K}ondo effect for a two-channel spin 
{S}=1
 impurity},
   volume={100},
     pages = {075434},
  numpages = {14},
   ISSN={2469-9969},
   url={http://dx.doi.org/10.1103/PhysRevB.100.075434},
   DOI={10.1103/physrevb.100.075434},
   number={7},
   journal={Phys. Rev. B},
   publisher={American Physical Society (APS)},
   author={Blesio, G. G. and Manuel, L. O. and Aligia, A. A. and Roura-Bas, P.},
   year={2019},
   month={Aug} }

@article{Blesio2018,
   title={Topological quantum phase transition between {F}ermi liquid phases in an {A}nderson impurity model},
   volume={98},
   ISSN={2469-9969},
   url={http://dx.doi.org/10.1103/PhysRevB.98.195435},
   DOI={10.1103/physrevb.98.195435},
   number={19},
     pages = {195435},
  numpages = {14},
   journal={Phys. Rev. B},
   publisher={American Physical Society (APS)},
   author={Blesio, G. G. and Manuel, L. O. and Roura-Bas, P. and Aligia, A. A.},
   year={2018},
   month={Nov} }

@article{Barral2017,
   title={Kondo behavior and conductance through 3d impurities in gold chains doped with oxygen},
   volume={146},
   ISSN={1089-7690},
   url={http://dx.doi.org/10.1063/1.4973982},
   DOI={10.1063/1.4973982},
   number={9},
   journal={J. Chem. Phys.},
   publisher={AIP Publishing},
   author={Barral, M. A. and Di Napoli, S. and Blesio, G. and Roura-Bas, P. and Camjayi, A. and Manuel, L. O. and Aligia, A. A.},
   year={2017},
   month={Mar},
   pages={092315} }

@article{KONDO-PAPER,
    author = {Kondo, Jun},
    title = {Resistance minimum in dilute magnetic alloys},
    journal = {Prog. Theor. Phys.},
    volume = {32},
    number = {1},
    pages = {37-49},
    year = {1964},
    month = {Jul},
    issn = {0033-068X},
    doi = {10.1143/PTP.32.37},
    url = {https://doi.org/10.1143/PTP.32.37},
}

@article{GAZZA-TORIO-2006,
   title={Anderson impurity in the one-dimensional {H}ubbard model for finite-size systems},
   volume={74},
   ISSN={1550-235X},
   url={http://dx.doi.org/10.1103/PhysRevB.74.195103},
   DOI={10.1103/physrevb.74.195103},
   number={19},
   pages = {195103},
  numpages = {9},
   journal={Phys. Rev. B},
   publisher={American Physical Society (APS)},
   author={Costamagna, S. and Gazza, C. J. and Torio, M. E. and Riera, J. A.},
   year={2006},
   month={Nov} }

@phdthesis{GERMAN-TESIS,
  title  = {F{\'\i}sica de {K}ondo y aproximaciones diagram{\'a}ticas},
  author = {Blesio, Germ{\'a}n G.},
  school = {Universidad Nacional de Rosario}, year = {2019}}

@phdthesis{HAND-TESIS,
  title  = {Finite-Size Effects in a One-Dimensional {K}ondo Box},
  author = {Hand, Thomas},
  school = {Universit{\"a}t Bonn}, year = {2006}}

@article{WU-2004,
   title={Quantum Phase Transitions and Bipartite Entanglement},
   volume={93},
   ISSN={1079-7114},
   url={http://dx.doi.org/10.1103/PhysRevLett.93.250404},
   DOI={10.1103/physrevlett.93.250404},
   number={25},
     pages = {250404},
  numpages = {4},
   journal={Phys. Rev. Lett.},
   publisher={American Physical Society (APS)},
   author={Wu, L.-A. and Sarandy, M. S. and Lidar, D. A.},
   year={2004},
   month={Dec} }

@article{DMRG-WHITE,
  title = {Density matrix formulation for quantum renormalization groups},
  author = {White, Steven R.},
  journal = {Phys. Rev. Lett.},
  volume = {69},
  issue = {19},
  pages = {2863--2866},
  numpages = {0},
  year = {1992},
  month = {Nov},
  publisher = {American Physical Society},
  doi = {10.1103/PhysRevLett.69.2863},
  url = {https://link.aps.org/doi/10.1103/PhysRevLett.69.2863}
}

@article{BORDA-2007,
  title = {Kondo screening cloud in a one-dimensional wire: Numerical renormalization group study},
  author = {Borda, L\'aszl\'o},
  journal = {Phys. Rev. B},
  volume = {75},
  pages = {041307(R)},
  year = {2007},
  doi = {10.1103/PhysRevB.75.041307},
}

@article{Borzenets2020,
    author={Borzenets, Ivan V.
    and Shim, Jeongmin
    and Chen, Jason C. H.
    and Ludwig, Arne
    and Wieck, Andreas D.
    and Tarucha, Seigo
    and Sim, H.-S.
    and Yamamoto, Michihisa},
    title={Observation of the {K}ondo screening cloud},
    journal={Nature},
    year={2020},
    month={Mar},
    day={01},
    volume={579},
    number={7798},
    pages={210-213},
        issn={1476-4687},
    doi={10.1038/s41586-020-2058-6},
    url={https://doi.org/10.1038/s41586-020-2058-6}
}

@article{CHO-17,
    author={Cho, Jaeyoon
    and Kim, Kun Woo},
    title={Quantum Phase Transition and Entanglement in Topological Quantum Wires},
    journal={Sci. Rep.},
    year={2017},
    month={Jun},
    day={05},
    volume={7},
    number={1},
    pages={2745},
    issn={2045-2322},
    doi={10.1038/s41598-017-02717-w},
    url={https://doi.org/10.1038/s41598-017-02717-w}
}

@inbook{Affleck10,
author = { Ian   Affleck },
title = {The {K}ondo Screening Cloud: What It Is and How to Observe It},
booktitle = {Perspectives of Mesoscopic Physics},
chapter = {},
pages = {1-44},
year = {2010},
publisher = {World Scientific},
doi = {10.1142/9789814299442_0001},
URL = {https://www.worldscientific.com/doi/abs/10.1142/9789814299442_0001},
}

@article{Sorensen2007,
  title   = {Impurity entanglement entropy and the {K}ondo screening cloud},
  author  = {S\o{}rensen, Erik S. and Chang, Ming-Shyang and Laflorencie, Nicolas and Affleck, Ian},
  journal = {J. Stat. Mech.},
  volume = {2007},
  pages   = {L01001},
  year    = {2007}, doi = {10.1088/1742-5468/2007/01/L01001}
}

@article{COX-ZAWADOWSKI-1998,
  title = {Exotic {K}ondo effects in metals: Magnetic ions in a crystalline electric field and tunnelling centres},
  author = {Cox, D. L. and Zawadowski, A.},
  journal = {Adv. Phys.},
  volume = {47},
  pages = {599},
  year = {1998},
  doi = {10.1080/000187398243500},
}

@article{Affleck2009,
  title   = {Entanglement entropy in quantum impurity systems and systems with boundaries},
  author  = {Affleck, Ian and Laflorencie, Nicolas and S\o{}rensen, Erik S.},
  journal = {J. Phys. A: Math. Theor.},
  volume  = {42},
  pages   = {504009},
  year    = {2009}
}

@article{Sorensen96,
  title = {Scaling theory of the {K}ondo screening cloud},
  author = {S\o{}rensen, Erik S. and Affleck, Ian},
  journal = {Phys. Rev. B},
  volume = {53},
  issue = {14},
  pages = {9153--9167},
  numpages = {0},
  year = {1996},
  month = {Apr},
  publisher = {American Physical Society},
  doi = {10.1103/PhysRevB.53.9153},
  url = {https://link.aps.org/doi/10.1103/PhysRevB.53.9153}
}

@article{Hand06,
  title = {Spin Correlations and Finite-Size Effects in the One-Dimensional {K}ondo Box},
  author = {Hand, Thomas and Kroha, Johann and Monien, Hartmut},
  journal = {Phys. Rev. Lett.},
  volume = {97},
  issue = {13},
  pages = {136604},
  numpages = {4},
  year = {2006},
  month = {Sep},
  publisher = {American Physical Society},
  doi = {10.1103/PhysRevLett.97.136604},
  url = {https://link.aps.org/doi/10.1103/PhysRevLett.97.136604}
}

@article{Busser10,
  title = {Numerical analysis of the spatial range of the {K}ondo effect},
  author = {B\"usser, C. A. and Martins, G. B. and Costa Ribeiro, L. and Vernek, E. and Anda, E. V. and Dagotto, E.},
  journal = {Phys. Rev. B},
  volume = {81},
  issue = {4},
  pages = {045111},
  numpages = {13},
  year = {2010},
  month = {Jan},
  publisher = {American Physical Society},
  doi = {10.1103/PhysRevB.81.045111},
  url = {https://link.aps.org/doi/10.1103/PhysRevB.81.045111}
}

@article{Ribeiro19,
  title = {Numerical study of the {K}ondo cloud using finite-{U} slave bosons},
  author = {Ribeiro, L. C. and Martins, G. B. and G\'omez-Silva, G. and Anda, E. V.},
  journal = {Phys. Rev. B},
  volume = {99},
  issue = {8},
  pages = {085139},
  numpages = {13},
  year = {2019},
  month = {Feb},
  publisher = {American Physical Society},
  doi = {10.1103/PhysRevB.99.085139},
  url = {https://link.aps.org/doi/10.1103/PhysRevB.99.085139}
}

@article{BARZYKIN-1996,
  title = {The {K}ondo screening cloud: What can we learn from perturbation theory?},
  author = {Barzykin, Victor and Affleck, Ian},
  journal = {Phys. Rev. Lett.},
  volume = {76},
  pages = {4959},
  year = {1996},
  doi = {10.1103/PhysRevLett.76.4959},
}

@article{BARZYNKIN-1998,
  title = {Screening cloud in the $k$-channel {K}ondo model: Perturbative and large-$k$ results},
  author = {Barzykin, Victor and Affleck, Ian},
  journal = {Phys. Rev. B},
  volume = {57},
  issue = {1},
  pages = {432--448},
  numpages = {0},
  year = {1998},
  month = {Jan},
  publisher = {American Physical Society},
  doi = {10.1103/PhysRevB.57.432},
  url = {https://link.aps.org/doi/10.1103/PhysRevB.57.432}
}

@article{NOZIERES,
	author = {{Ph. Nozi\`eres} and {A. Blandin}},
	title = {Kondo effect in real metals},
	DOI= "10.1051/jphys:01980004103019300",
	url= "https://doi.org/10.1051/jphys:01980004103019300",
	journal = {J. Phys. France},
	year = 1980,
	volume = 41,
	number = 3,
	pages = "193-211",
}

@article{BLESIO-ALIGIA-2023,
  title = {Topological quantum phase transition in individual {F}e atoms on {MoS}$_2$/{A}u(111)},
  author = {Blesio, G. G. and Aligia, A. A.},
  journal = {Phys. Rev. B},
  volume = {108},
  pages = {045113},
  year = {2023},
  doi = {10.1103/PhysRevB.108.045113}
}

@article{BLESIO-SCIPOST-2023,
  title = {Topological quantum phase transition of nickelocene on {C}u(100)},
  author = {Blesio, Germ{\'a}n G. and {\v Z}itko, Rok and Manuel, Luis O. and Aligia, Armando A.},
  journal = {SciPost Phys.},
  volume = {14},
  pages = {042},
  year = {2023},
  doi = {10.21468/SciPostPhys.14.3.042}
}

@article{BLESIO-REVIEW-2024,
  title = {Anisotropy-driven topological quantum phase transition in magnetic impurities},
  author = {Blesio, G. G. and Manuel, L. O. and Aligia, A. A.},
  journal = {Mater. Quantum Technol.},
  volume = {4},
  pages = {042001},
  year = {2024}
}

@article{Nevidomskyy2009,
  title = {Kondo Resonance Narrowing in $d$- and $f$-Electron Systems},
  author = {Nevidomskyy, Andriy H. and Coleman, P.},
  journal = {Phys. Rev. Lett.},
  volume = {103},
  issue = {14},
  pages = {147205},
  numpages = {4},
  year = {2009},
  month = {Oct},
  publisher = {American Physical Society},
  doi = {10.1103/PhysRevLett.103.147205},
  url = {https://link.aps.org/doi/10.1103/PhysRevLett.103.147205}
}

@article{DEBERTOLIS2021,
  title = {Few-body nature of {K}ondo correlated ground states},
  author = {Debertolis, Maxime and Florens, Serge and Snyman, Izak},
  journal = {Phys. Rev. B},
  volume = {103},
  issue = {23},
  pages = {235166},
  numpages = {11},
  year = {2021},
  month = {Jun},
  publisher = {American Physical Society},
  doi = {10.1103/PhysRevB.103.235166},
  url = {https://link.aps.org/doi/10.1103/PhysRevB.103.235166}
}

@article{Orlandini2025,
  title = {Nontrivial entanglement passively mediated by a quenched magnetic impurity},
  author = {Orlandini, A. A. and Blesio, G. G. and Gazza, C. J. and Manuel, L. O.},
  journal = {Phys. Rev. B},
  volume = {112},
  issue = {8},
  pages = {L081403},
  numpages = {6},
  year = {2025},
  month = {Aug},
  publisher = {American Physical Society},
  doi = {10.1103/kx2n-8rn8},
  url = {https://link.aps.org/doi/10.1103/kx2n-8rn8}
}

@article{Kouwenhoven2001,
  author  = {Kouwenhoven, L. and Glazman, L.},
  title   = {Revival of the {K}ondo effect},
  journal = {Physics World},
  volume  = {14},
  pages   = {33},
  year    = {2001}
}

@article{Georges1996,
  author  = {Georges, A. and Kotliar, G. and Krauth, W. and Rozenberg, M. J.},
  title   = {Dynamical mean-field theory of strongly correlated fermion systems and the limit of infinite dimensions},
  journal = {Rev. Mod. Phys.},
  volume  = {68},
  pages   = {13},
  year    = {1996},
  doi     = {10.1103/RevModPhys.68.13}
}

@article{Liang2002,
  author  = {Liang, W. and Shores, M. P. and Bockrath, M. and Long, J. R. and Park, H.},
  title   = {Kondo resonance in a single-molecule transistor},
  journal = {Nature},
  volume  = {417},
  pages   = {725},
  year    = {2002},
  doi     = {10.1038/nature00790}
}

@article{Parks2010,
  author  = {Parks, J. J. and Champagne, A. R. and Costi, T. A. and Shum, W. W. and Pasupathy, A. N. and Neuscamman, E. and Flores-Torres, S. and Cornaglia, P. S. and Aligia, A. A. and Balseiro, C. A. and Chan, G. K.-L. and Abru{\~n}a, H. D. and Ralph, D. C.},
  title   = {Mechanical Control of Spin States in Spin-1 Molecules and the Underscreened {K}ondo Effect},
  journal = {Science},
  volume  = {328},
  pages   = {1370},
  year    = {2010},
  doi     = {10.1126/science.1186874}
}

@article{GoldhaberGordon1998,
  author  = {Goldhaber-Gordon, D. and Shtrikman, H. and Mahalu, D. and Abusch-Magder, D. and Meirav, U. and Kastner, M. A.},
  title   = {Kondo effect in a single-electron transistor},
  journal = {Nature},
  volume  = {391},
  pages   = {156},
  year    = {1998},
  doi     = {10.1038/34373}
}

@article{Cronenwett1998,
  author  = {Cronenwett, S. M. and Oosterkamp, T. H. and Kouwenhoven, L. P.},
  title   = {A Tunable {K}ondo Effect in Quantum Dots},
  journal = {Science},
  volume  = {281},
  pages   = {540},
  year    = {1998},
  doi     = {10.1126/science.281.5376.540}
}

@article{Pustilnik2001,
  author  = {Pustilnik, M. and Glazman, L. I.},
  title   = {Kondo Effect in Real Quantum Dots},
  journal = {Phys. Rev. Lett.},
  volume  = {87},
  pages   = {216601},
  year    = {2001},
  doi     = {10.1103/PhysRevLett.87.216601}
}

@article{Potok2007,
  author  = {Potok, R. M. and Rau, I. G. and Shtrikman, H. and Oreg, Y. and Goldhaber-Gordon, D.},
  title   = {Observation of the two-channel {K}ondo effect},
  journal = {Nature},
  volume  = {446},
  pages   = {167},
  year    = {2007},
  doi     = {10.1038/nature05556}
}

@article{Iftikhar2015,
  author  = {Iftikhar, Z. and Jezouin, S. and Anthore, A. and Gennser, U. and Parmentier, F. D. and Cavanna, A. and Pierre, F.},
  title   = {Two-channel {K}ondo effect and renormalization flow with macroscopic quantum charge states},
  journal = {Nature},
  volume  = {526},
  pages   = {233},
  year    = {2015},
  doi     = {10.1038/nature15384}
}

@article{Madhavan1998,
  author  = {Madhavan, V. and Chen, W. and Jamneala, T. and Crommie, M. F. and Wingreen, N. S.},
  title   = {Tunneling into a Single Magnetic Atom: Spectroscopic Evidence of the {K}ondo Resonance},
  journal = {Science},
  volume  = {280},
  pages   = {567},
  year    = {1998},
  doi     = {10.1126/science.280.5363.567}
}

@article{Li1998,
  author  = {Li, J. and Schneider, W.-D. and Berndt, R. and Delley, B.},
  title   = {Kondo Scattering Observed at a Single Magnetic Impurity},
  journal = {Phys. Rev. Lett.},
  volume  = {80},
  pages   = {2893},
  year    = {1998},
  doi     = {10.1103/PhysRevLett.80.2893}
}

@article{Otte2008,
  author  = {Otte, A. F. and Ternes, M. and von Bergmann, K. and Loth, S. and Brune, H. and Lutz, C. P. and Hirjibehedin, C. F. and Heinrich, A. J.},
  title   = {The role of magnetic anisotropy in the {K}ondo effect},
  journal = {Nat. Phys.},
  volume  = {4},
  pages   = {847},
  year    = {2008},
  doi     = {10.1038/nphys1072}
}

@article{Minamitani2012,
  author  = {Minamitani, E. and Tsukahara, N. and Matsunaka, D. and Kim, Y. and Takagi, N. and Kawai, M.},
  title   = {Symmetry-Driven Novel {K}ondo Effect in a Molecule},
  journal = {Phys. Rev. Lett.},
  volume  = {109},
  pages   = {086602},
  year    = {2012},
  doi     = {10.1103/PhysRevLett.109.086602}
}

@Article{Spinelli2015,
author={Spinelli, A.
and Gerrits, M.
and Toskovic, R.
and Bryant, B.
and Ternes, M.
and Otte, A. F.},
title={Exploring the phase diagram of the two-impurity {K}ondo problem},
journal={Nat. Commun.},
year={2015},
month={Nov},
day={30},
volume={6},
number={1},
pages={10046},
issn={2041-1723},
doi={10.1038/ncomms10046},
url={https://doi.org/10.1038/ncomms10046}
}

@article{Ormaza2017,
  author  = {Ormaza, M. and Abufager, P. and Verlhac, B. and Bachellier, N. and Bocquet, M.-L. and Lorente, N. and Limot, L.},
  title   = {Controlled spin switching in a metallocene molecular junction},
  journal = {Nat. Commun.},
  volume  = {8},
  pages   = {1974},
  year    = {2017},
  doi     = {10.1038/s41467-017-02151-6}
}

@article{Hiraoka2017,
  author  = {Hiraoka, R. and Minamitani, E. and Arafune, R. and Tsukahara, N. and Watanabe, S. and Kawai, M. and Takagi, N.},
  title   = {Single-molecule quantum dot as a {K}ondo simulator},
  journal = {Nat. Commun.},
  volume  = {8},
  pages   = {16012},
  year    = {2017},
  doi     = {10.1038/ncomms16012}
}

@article{Evers2020,
  author  = {Evers, F. and Koryt{\'a}r, R. and Tewari, S. and van Ruitenbeek, J. M.},
  title   = {Advances and challenges in single-molecule electron transport},
  journal = {Rev. Mod. Phys.},
  volume  = {92},
  pages   = {035001},
  year    = {2020},
  doi     = {10.1103/RevModPhys.92.035001}
}

@article{Yang2019,
  author  = {Yang, K. and Paul, W. and Phark, S.-H. and Willke, P. and Bae, Y. and Choi, T. and Esat, T. and Ardavan, A. and Heinrich, A. J. and Lutz, C. P.},
  title   = {Coherent spin manipulation of individual atoms on a surface},
  journal = {Science},
  volume  = {366},
  pages   = {509},
  year    = {2019},
  doi     = {10.1126/science.aay6779}
}

@article{Verlhac2019,
  author  = {Verlhac, B. and Bachellier, N. and Garnier, L. and Ormaza, M. and Abufager, P. and Robles, R. and Bocquet, M.-L. and Ternes, M. and Lorente, N. and Limot, L.},
  title   = {Atomic-scale spin sensing with a single molecule at the apex of a scanning tunneling microscope},
  journal = {Science},
  volume  = {366},
  pages   = {623},
  year    = {2019},
  doi     = {10.1126/science.aax8222}
}

@article{Nishikawa2025,
  title = {Quantum entanglement in a pure state of strongly correlated quantum impurity systems},
  author = {Nishikawa, Yunori and Yoshioka, Tomoki},
  journal = {Phys. Rev. B},
  volume = {111},
  pages = {035112},
  year = {2025},
  doi = {10.1103/PhysRevB.111.035112},
}

@article{Alkurtass2016,
  title = {Entanglement structure of the two-channel {K}ondo model},
  author = {Alkurtass, Bedoor and Bayat, Abolfazl and Affleck, Ian and Bose, Sougato and Johannesson, Henrik and Sodano, Pasquale and S\o{}rensen, Erik S. and Le Hur, Karyn},
  journal = {Phys. Rev. B},
  volume = {93},
  pages = {081106(R)},
  year = {2016},
  doi = {10.1103/PhysRevB.93.081106},
}

@article{Zhou2026,
  title = {Universality of energy-space entanglement in quantum impurity models},
  author = {Zhou, Geng-Dong and Song, Zhi-Da},
  journal = {arXiv:2607.27394},
  year = {2026},
}

@article{Stocker2026,
  title = {Unraveling screening mechanisms in {K}ondo impurities using an {NRG-MPS}-based method},
  author = {Stocker, Lidia and Zilberberg, Oded},
  journal = {Phys. Rev. B},
  volume = {113},
  pages = {205104},
  year = {2026},
  doi = {10.1103/f88w-hg3j},
}

@article{Kim2026,
  title = {Universal spin screening clouds in local moment phases},
  author = {Kim, Minsoo L. and Shim, Jeongmin and Sim, H.-S. and Kim, Donghoon},
  journal = {Phys. Rev. B},
  volume = {113},
  pages = {L241401},
  year = {2026},
  doi = {10.1103/dxvl-kdlw},
}

@article{Sorensen2024,
  title = {Kondo screening cloud scaling: impurity entanglement and magnetization},
  author = {S\o{}rensen, Erik S.},
  journal = {J. Stat. Mech.},
  volume = {2024},
  pages = {013106},
  year = {2024},
  doi = {10.1088/1742-5468/ad1d53},
}

@article{CardenesWuttig2026,
  title = {Probing impurity quantum criticality with entanglement witnesses},
  author = {C{\'a}rdenes Wuttig, Mateo and Millis, Andrew J.},
  journal = {arXiv:2608.12317},
  year = {2026},
}

@article{Gaines2026,
  title = {Spin-chain multichannel {K}ondo model via image impurity boundary condition},
  author = {Gaines, Jordan and Li, Guangjie and V\"ayrynen, Jukka I.},
  journal = {Phys. Rev. Lett.},
  volume = {136},
  pages = {036503},
  year = {2026},
  doi = {10.1103/wj75-tg3z},
}

@article{Gubernatis1987,
  title = {Spin and charge correlations around an {A}nderson magnetic impurity},
  author = {Gubernatis, J. E. and Hirsch, J. E. and Scalapino, D. J.},
  journal = {Phys. Rev. B},
  volume = {35},
  pages = {8478},
  year = {1987},
  doi = {10.1103/PhysRevB.35.8478},
}

\appendix

\section{Integrated correlation function for the 2CAM} 
\label{App:Sigma}

Let $\mathbf{S}_{tot}$ denote the total spin of the system, $\mathbf{S}_{tot}=\mathbf{S}_{imp}+\sum_{i,\alpha}\mathbf{s}_{i\alpha}$. Here, $\mathbf{S}_{imp}=\mathbf{S}_L+\mathbf{S}_R$ is the spin of the two-orbital impurity and the index $i$ and $\alpha=L,R$ covers the sites of both chains [see Fig.~\ref{fig: modelo}(a)]. Accordingly, the correlation between $\mathbf{S}_{tot}$ and the spin of the left orbital of the impurity, $\mathbf{S}_L$, splits into three contributions,
\begin{equation}
    \langle\mathbf{S}_L\cdot \mathbf{S}_{tot}\rangle=\langle\mathbf{S}_L\cdot\mathbf{S}_L\rangle+\langle\mathbf{S}_L\cdot\mathbf{S}_R\rangle+\langle\mathbf{S}_L\cdot\sum\mathbf{s}_{i\alpha}\rangle.
    \label{ec: SMproduct}
\end{equation}

For $D=0$ in the Kondo phase the ground state is a spin singlet. Hence $\mathbf{S}_{tot}|\Psi\rangle = 0$, and the left-hand side of Eq.~(\ref{ec: SMproduct}) is zero. Rearranging the remaining terms yields
\begin{equation}
    1+\frac{\sum\langle\mathbf{S}_L\cdot\mathbf{s}_{i\alpha}\rangle}{\langle\mathbf{S}_L\cdot\mathbf{S}_L\rangle+\langle\mathbf{S}_L\cdot\mathbf{S}_R\rangle}=0,
\end{equation}
that is satisfied only when the sum extends over every site of the two conduction chains. Restricting the sum to the first $x$ sites of each chain allows us to introduce a function $\Sigma(x)$ that measures the screening of the impurity
\begin{equation}
    \Sigma(x)=1+\sum_{i=1}^x \sum_{\alpha=L,R}\frac{\langle \mathbf{S}_{L}\cdot\mathbf{s}_{i\alpha}\rangle}{\langle\mathbf{S}_{L}\cdot\mathbf{S}_{L}\rangle + \langle\mathbf{S}_{L}\cdot\mathbf{S}_{R}\rangle},
    \label{ec: SMsigma_function}
\end{equation}
which is Eq.~(\ref{ec: sigma_function}) of the main text.
In the last equation, $\Sigma(x=0)=1$ corresponds to the case where no chain sites are taken, and therefore the second term in the equation vanishes. 
Due to symmetry, Eq. (\ref{ec: SMsigma_function}) can be equivalently defined with either $\mathbf{S}_L$ or $\mathbf{S}_R$ in the second term.
For $D\neq0$ total spin is not conserved and the sum rule above no longer holds.

\section{Toy model for the Anderson model} 
\label{App:toy}

The spin-1 Anderson impurity model coupled to two conduction baths can be simplified by 
replacing each bath, as well as each orbital of the impurity, with a single spin-$1/2$ 
degree of freedom. The orbitals $a$ and $b$ of the impurity, replaced by the spin 
operators $\hat{\mathbf{S}}_a$ and $\hat{\mathbf{S}}_b$, respectively, are subject to a strong Hund's coupling $J_H$  yielding a total spin $S=1$. 
The impurity spin operators $\hat{\mathbf{S}}_a$ and $\hat{\mathbf{S}}_b$  are coupled via  exchange interactions $J_R$ and $J_L$ to spin-1/2 operators $\hat{\mathbf{s}}_R$ and $\hat{\mathbf{s}}_L$, representing the right and left conduction electrons, respectively. Additionally, the impurity has a single-ion anisotropy characterized by a parameter $D$. So the Hamiltonian of the resulting effective model takes the form:
\begin{equation}
    \begin{split}
        \hat{H} =& J_L \hat{\mathbf{s}}_L \cdot \hat{\mathbf{S}}_a + J_R \hat{\mathbf{s}}_R \cdot \hat{\mathbf{S}}_b - J_H \hat{\mathbf{S}}_a \cdot \hat{\mathbf{S}}_b \\
        &+ D\left(\hat{S}^z_a+\hat{S}^z_b\right)^2,
    \end{split}
\end{equation} 
which reduces to Eq.~(\ref{ec: toy}) of the main text for $J_L=J_R=J$.

For $J_H\to\infty$ the impurity is a pure spin~1. Projecting onto its triplet, $P\hat{\mathbf S}_{a,b}P=\hat{\mathbf S}_{imp}/2$, with $P$ the projector onto the triplet and $\hat{\mathbf S}_{imp}=\hat{\mathbf S}_a+\hat{\mathbf S}_b$, and taking $J_L=J_R=J$, the Hamiltonian becomes
\begin{equation*}
    \hat H_{J_H\to\infty}=\tilde J\,\hat{\mathbf S}_{imp}\cdot\hat{\mathbf S}_{ch}+D\,(\hat S^z_{imp})^2,\qquad \tilde J=J/2,
\end{equation*}
with $\hat{\mathbf S}_{ch}=\hat{\mathbf s}_L+\hat{\mathbf s}_R$. The sector with $S_{ch}=0$ decouples from the impurity and always lies higher in energy, so the two conduction spins, which never interact directly, form a spin~1 and $\langle\hat{\mathbf s}_L\cdot\hat{\mathbf s}_R\rangle=1/4$ for every $D$. In the sector with $\hat S^z_{imp}+\hat S^z_{ch}=0$ the ground state is
\begin{equation*}
    |\psi\rangle=\cos\theta\,|0,0\rangle-\sin\theta\,\frac{|1,\bar1\rangle+|\bar1,1\rangle}{\sqrt2},\qquad
    \tan2\theta=\frac{2\sqrt2\,\tilde J}{D-\tilde J},
\end{equation*}
with $0\le2\theta<\pi$, states labeled $|M_{imp},M_{ch}\rangle$ by the eigenvalues of $\hat S^z_{imp}$ and $\hat S^z_{ch}$, and $\bar1\equiv-1$. It interpolates between $(|1,\bar1\rangle+|\bar1,1\rangle)/\sqrt2$ for $D\ll-\tilde J$, which is not an SU(2) singlet but a superposition of total spin $0$ and $2$, the isotropic singlet $(|1,\bar1\rangle-|0,0\rangle+|\bar1,1\rangle)/\sqrt3$ at $D=0$, and the product state $|0,0\rangle$ for $D\gg\tilde J$. Correspondingly, the correlation between each conduction spin and the impurity orbital it is coupled to takes the values $-1/4$, $-1/2$, and $0$, as in Fig.~\ref{fig: natural_orbitals_correlations}(c). The ground state is the lowest eigenvector of a $2\times2$ block with off-diagonal element $\sqrt2\tilde J\neq0$; it is nondegenerate and analytic in $D$, so the toy model has a crossover at $D\sim\tilde J$ but no transition.

Since $|\psi\rangle$ has $\hat S^z_{imp}+\hat S^z_{ch}=0$ and is invariant under a global spin flip, each conduction spin is maximally mixed, and the channel cut gives $\ln2$ for every $D$. The impurity cut follows Eq.~(\ref{ec:Simpq}) with $q=\sin^2\theta$. The symmetric cut gives $\ln2$ for $D\ll-\tilde J$, $\ln(4/\sqrt3)\simeq0.84$ at $D=0$, and $\ln4$ for $D\gg\tilde J$, where $|0,0\rangle=|T_0\rangle_{imp}|T_0\rangle_{ch}$ and the cut splits both triplets, each contributing $\ln2$.

\section{Effective exchange between the channels}
\label{App:Heff}

For $D$ and $J_H$ much larger than $J$ and than the bandwidth, take $V_J=J\sum_\alpha\hat{\mathbf S}_\alpha\cdot\hat{\mathbf s}_{1\alpha}$ as the perturbation of the impurity ground state $|T_0\rangle$, with $|T_\pm\rangle$ at energy $D$ and the singlet $|S_0\rangle$ at energy $J_H$. Using $\hat S^\pm_L|T_0\rangle=\hat S^\pm_R|T_0\rangle=|T_\pm\rangle/\sqrt2$ and $\hat S^z_L|T_0\rangle=-\hat S^z_R|T_0\rangle=|S_0\rangle/2$, second order gives
\begin{align}
H_{\rm eff}&=-\frac{J^2}{8D}\sum_{\pm}\big(\hat s^\mp_{1L}+\hat s^\mp_{1R}\big)\big(\hat s^\pm_{1L}+\hat s^\pm_{1R}\big)
-\frac{J^2}{4J_H}\big(\hat s^z_{1L}-\hat s^z_{1R}\big)^2,\nonumber
\end{align}
which equals Eq.~(\ref{ec:Heff}) up to a constant. For two spins $1/2$ its ground state is $|T_0\rangle_{ch}$, with gaps $J^2/4D+J^2/4J_H$ to $|T_\pm\rangle_{ch}$ and $J^2/2D$ to $|S_0\rangle_{ch}$.

\end{document}